\documentclass[]{gji}

\usepackage{timet}
\usepackage{graphicx}
\usepackage{color}
\usepackage{acronym}
\let\bibhang\relax
\usepackage{natbib}
\usepackage{lineno}
\newacronym{GPS}{global positioning system}[GPS]
\newacronym{FFT}{fast Fourier transform}[FFT]
\newacronym{BGI}{Bureau Gravim$\acute{\rm e}$trique International}[BGI98]
\newacronym{NKG}{Nordic Geodetic Commission}[NKG]
\newacronym{OCTAS}{Ocean Circulation and Transport Between North Atlantic and the Arctic Sea}[OCTAS]
\newacronym{GOCINA}{Geoid and Ocean Circulation in the North Atlantic}[GOCINA]
\newacronym{OCCAM}{Ocean Circulation and Climate Advanced Modelling}[OCCAM]
\newacronym{NIMA}{National Imagery and Mapping Agency}[NIMA]
\newacronym{NMCA}{Norwegian Mapping and Cadastre Authority}[NMCA]
\newacronym{NMA}{Norwegian Mapping Authority}[NMA]
\newacronym{IGN}{Institut Geographique National}[IGN]
\newacronym{IGS}{International GNSS Service}[IGS]
\newacronym{IVS}{International VLBI Service}[IVS]
\newacronym{IDS}{International DORIS Service}[IDS]
\newacronym{ITRS}{International Terrestrial Reference System}[ITRF]
\newacronym{ITRF}{International Terrestrial Reference Frame}[ITRF2014]
\newacronym{IERS}{International Earth Rotation Service}[IERS]
\newacronym{GNSS}{Global Navigation Satellite System}[GNSS]
\newacronym{VLBI}{Very Long Baseline Interfer}[VLBI]
\newacronym{CGPS}{Continuous GPS}[CGPS]
\newacronym{DGPS}{Differential GPS}[DGPS]
\newacronym{DoD}{Department of Defence}[DoD]
\newacronym{DORIS}{Doppler Orbitography and Radiopositioning  Integrated by Satellites}[DORIS]
\newacronym{EGM}{Earth Gravity Model}[EGM]
\newacronym{EGNOS}{European Geostationary Navigation Overlay Service}[EGNOS]
\newacronym{EOP}{Earth Orientation Parameters}[EOP]
\newacronym{ESA}{European Space Agency}[ESA]
\newacronym{ESOC}{European Space Operation Centre}[ESOC]
\newacronym{ESEAS}{European Sea Level Service}[ESEAS]
\newacronym{ESEAS-RI}{ESEAS Research Infrastructure}[ESEAS-RI]
\newacronym{ETRS}{European Terrestrial Reference System}[ETRS]
\newacronym{EUREF}{IAG Reference Frame Sub-Commission for Europe}[EUREF] 
\newacronym{EPN}{EUREF Permanent Network}[EPN]
\newacronym{NGS}{National Geodetic Survey}[NGS]
\newacronym{GAMIT}{GPS Analysis Software of MIT}[GAMIT]
\newacronym{GIPSY}{GNSS-Inferred Positioning System}[GIPSY]
\newacronym{GLONASS}{Global Navigation Satellite System}[GLONAS]
\newacronym{GPS}{Global Positioning System}[GPS]
\newacronym{IERS}{International Earth Rotation Service}[IERS]
\newacronym{IGS-TIGA}{IGS-Tide Gauge Benchmark Monitoring }[IGS-TIGA]
\newacronym{JPL}{Jet Propulsion Laboratory}[JPL]
\newacronym{UNR}{University of Nevada, Reno}[UNR]
\newacronym{MIT}{Massachusetts Institute of Technology}[MIT]
\newacronym{NIMA}{National Imagery and Mapping Agency}[NIMA]
\newacronym{NNSS}{Navy Navigation satellite system}[NNSS]
\newacronym{NLES}{Navigation Land Earth Station}[NLES]
\newacronym{NASA}{National Aeronautics \& Space Administration}[NASA]
\newacronym{OSO}{Onsala Space Observatory}[OSO]
\newacronym{RIMS}{Ranging and Integrity Monitoring Station}[RIMS]
\newacronym{RINEX}{Receiver independent exchange format}[RINEX]
\newacronym{SINEX}{Site independent exchange format}[SINEX]
\newacronym{SLR}{Satellite Laser Ranging}[SLR]
\newacronym{SOPAC}{Scripps Orbits and Permanent Array Center}[SOPAC]
\newacronym{SWEPOS}{Swedish GPS Network}[SWEPOS]
\newacronym{SATREF}{Satellite based reference system}[SATREF]
\newacronym{VLBI}{Very Long Baseline Interferometry}[VLBI]
\newacronym{WGS84}{World Geodetic System 1984}[WGS84]
\newacronym{GFZ}{GeoForschungsZentrum Potsdam}[GFZ]
\newacronym{GIG}{GPS IERS and Geodynamics 1991}[GIG91]
\newacronym{PPP}{Precise Point Positioning}[PPP]
\newacronym{CHAMP}{CHAllenging Minisatellite Payload}[CHAMP]
\newacronym{AS}{Antispoofing}[AS]
\newacronym{GRACE}{Gravity Recovery and Climate Experiment}[GRACE]
\newacronym{PGR}{Post Glacial Rebound}[PGR]
\newacronym{PDIM}{Present Day Ice Melt}[PDIM]
\newacronym{TGB}{Tide Gauge Benchmar}[TGB]
\newacronym{SCG}{Super Conducting Gravity}[SCG]
\newacronym{GGOS}{Global Geodetic Observing System}[GGOS]
\newacronym{DD}{Double Difference}[DD]
\newacronym{IPEV}{French Polar Institute - Paul Emile Victor}[IPEV]
\newacronym{LSQ}{Least Square}[LS]
\newacronym{PSMSL}{Permanent Service for Mean Sea Level}[PSMSL]
\newacronym{PCV}{Phase Center Variation}[PCV]
\newacronym{GIA}{Glacial Isostatic Adjustment}[GIA]
\newacronym{GLOBK}{Global Kalman filter VLBI and GPS analysis program}[GLOBK]
\newacronym{BIFROST}{Baseline Inferences for Fennoscandian Rebound, Sea-level, and Tectonics}[BIFROST]
\newacronym{QIF}{Quasi Ionosphere Free}[QIF]
\newacronym{VMF}{Vienna Mapping Functions}[VMF1]
\newacronym{LGM}{Latest Glacial Maximum}[LGM]
\newacronym{SRIF}{Square root information filter}[SRIF]
\newacronym{CORS}{Continuously Operating Reference Station}[CORS]
\newacronym{epGNSS}{episodic GNSS stations}[epGNSS]
\newacronym{cGNSS}{continuous operating GNSS stations}[cGNSS]
\newacronym{RTK}{Real Time Kinematic}[RTK]
\newacronym{VGOS}{VLBI Global Observing System}[VGOS]
\newacronym{CMS}{Common Mode Signal}[CMS]
\newacronym{CM}{Common Mode}[CM]
\newacronym{LCM}{Load and Common Mode}[LCM]
\newacronym{EOF}{Empirical Ortoghonal Functions}[EOF]
\newacronym{LIA}{Little Ice Age}[LIA]
\newacronym{CMB}{Climatic Mass Balance}[CMB]
\newacronym{GS}{Glaciers and Snow}[GS]
\newacronym{ATM}{Atmospheric}[ATM]
\newacronym{NTO}{Non-Tidal Ocean}[NTO]
\newacronym{HYD}{Hydrological}[HYD]
\newacronym{AOH}{}[]
\newacronym{RMS}{Root Mean Square}[RMS]
\newacronym{LWS}{Land Water Storage}[LWS]
\newacronym{NTL}{Non-tidal loading}[NTL]
\newacronym{NNR}{No-Net-Rotation}[NNR]

\usepackage{hyperref}

\hypersetup{
             colorlinks=true,
             linkcolor=black,
             filecolor=black,
             citecolor=black,
             urlcolor=blue
           }

\date{Accepted 2021 November 24}
\pagerange{}
\volume{}
\pubyear{}

\usepackage{ulem}
\renewcommand{\em}{\it} 
\usepackage{xcolor}
\definecolor{Dred}{rgb}{0.312,0.070,0.070}
\definecolor{redc}{rgb}{0.999,0.000,0.000}

\newcounter{note}
\let\oldmarginpar\marginpar
\renewcommand\marginpar[1]{\-\oldmarginpar[\raggedleft\footnotesize #1]%
{\raggedright\footnotesize #1}}

\newcommand\strike{\bgroup\markoverwith{\textcolor{red}{\rule[0.5ex]{2pt}{1.0pt}}}\ULon}

\newcommand{\web}[1]{\url{#1}}

\title{Seasonal glacier and snow loading in Svalbard recovered from geodetic observations}
\author[H.P. Kierulf \& al.]{H.P. Kierulf$^{1,2}$,\thanks{halfdan.kierulf@kartverket.no} W.J.J. van Pelt$^3$, L. Petrov$^4$, M. D\"ahnn$^1$, A.-S. Kirkvik$^1$, O. Omang$^1$\\$^1$Geodetic Institute, Norwegian Mapping Authority, H{\o}nefoss, Norway.\\$^2$Department of Geosciences, University of Oslo, Oslo, Norway.\\$^3$Department of Earth Sciences, Uppsala University, Uppsala, Sweden.\\$^4$NASA Goddard Space Flight Center, Greenbelt, US}

\begin{document}
\label{firstpage}
\maketitle
\begin{summary}
  
  We processed time series from seven Global Navigation Satellite System (GNSS) stations and one Very Long Baseline Interferometry (VLBI) station in Svalbard. The goal was to capture the seasonal vertical displacements caused by elastic response of variable mass load due to ice and snow accumulation.
  We found that estimates of the annual signal in different GNSS solutions disagree by more than 3~mm which makes geophysical interpretation of raw GNSS time series problematic. To overcome this problem, we have used an enhanced Common Mode (CM) filtering technique. The time series are differentiated by the time series from remote station BJOS with known mass loading signals removed a priori.  Using this technique, we have achieved a substantial reduction of the differences between the GNSS solutions.
  We have computed mass loading time series from a regional
Climatic Mass Balance (CMB) and snow model that provides the amount of water equivalent at 
a 1~km resolution with a time step of 7~days.
We found that the entire 
vertical loading signal is present in data of two totally independent 
techniques at a statistically significant level of 95\%. This allowed us to conclude that the remaining 
errors in vertical signal derived from the CMB model are less 
than 0.2~mm at that significance level.
  Refining the land water storage loading model with a CMB model resulted in a reduction
of the annual amplitude from $2.1$~mm to $1.1$~mm in the CM filtered 
time series, while it had only a marginal impact on raw time series.
This provides a strong evidence that CM filtering is essential
for revealing local periodic signals when a millimetre level of 
accuracy is required.

\end{summary}



\section{Introduction}
\label{intro}
The Arctic archipelago Svalbard is exposed to climate change phenomena, the temperature is rising, the permafrost is melting, the sea level is rising, and the glaciers are retreating \citep[][]{hanssen-bauer++2019}. Consequences of climate change, like sea-level rise or increased land-uplift, can be observed by geodetic techniques in an accurate geodetic reference frame. On the other hand, these changes challenge the stability of the geodetic reference frame itself, e.g., the increased land uplift will deform the reference frame over time. Knowledge about the interaction between geophysical processes, crustal deformations and reference frame is mandatory to achieve the GGOS2020 goal of a reference frame with a stability of $0.1$~mm/yr \citep[][]{plag+pearlman2009}.

The geodetic observatory in Ny-\AA lesund is one of the core stations in the global geodetic network. It was established during the 1990s with \GNSS\ antennas, \VLBI\ telescope, \SCG, absolute gravity points, and control networks \citep[][]{kierulf++2009b}.

Due to Svalbard's remote location and challenging environmental conditions Ny-\AA lesund was for a long time the only location with permanent geodetic equipment on the archipelago. \cite{sato++2006,sato++06b} studied the gravity signal in Ny-\AA lesund and the interaction between gravity changes and uplift. The uplift in Ny-\AA lesund is not linear. It has a seasonal component, that will be studied in details in this manuscript, and an inter-annual signal induced by the long term (years to decades) evolution of glacier mass balance \cite[e.g.][]{kierulf++2009b}.   \cite{kierulf++2009a} showed that the uplift changed from year to year and that these variations are very well explained by the changes in the mass balance of the nearby glaciers. \cite{omang+kierulf2011} found that also the gravity rate is changing with time. \cite{memin++2012} showed that topography of glaciers has a significant effect on the gravity rate. The visco-elastic response of the last ice age \citep[][]{auriac++2016} and the visco-elastic response of the glacier retreat after the \LIA\ \citep{memin++2014} also contribute to the uplift in Ny-\AA lesund. In 2005, the Polish research station in Hornsund installed a new GNSS antenna. \cite{rajner2018} compared results from the stations in Hornsund and Ny-\AA lesund and demonstrated that both locations have non-linear uplift. All these papers focus mainly on glacier related phenomena with time spans ranging from years to decades or thousands of years.

The most prominent variations in snowpack and glacier mass are the annual cycle with accumulation of snow each winter and melting in the short Arctic summer. The crusts elastic response of this seasonal variations results in a seasonal cycle also in the GNSS station coordinates and other geodetic equipment. The crust is also exposed to \NTL\ from atmosphere, ocean and land water \citep[][]{petrov+boy2004,memin++2020}.

\begin{figure}
\centering
\includegraphics[width=.49\textwidth]{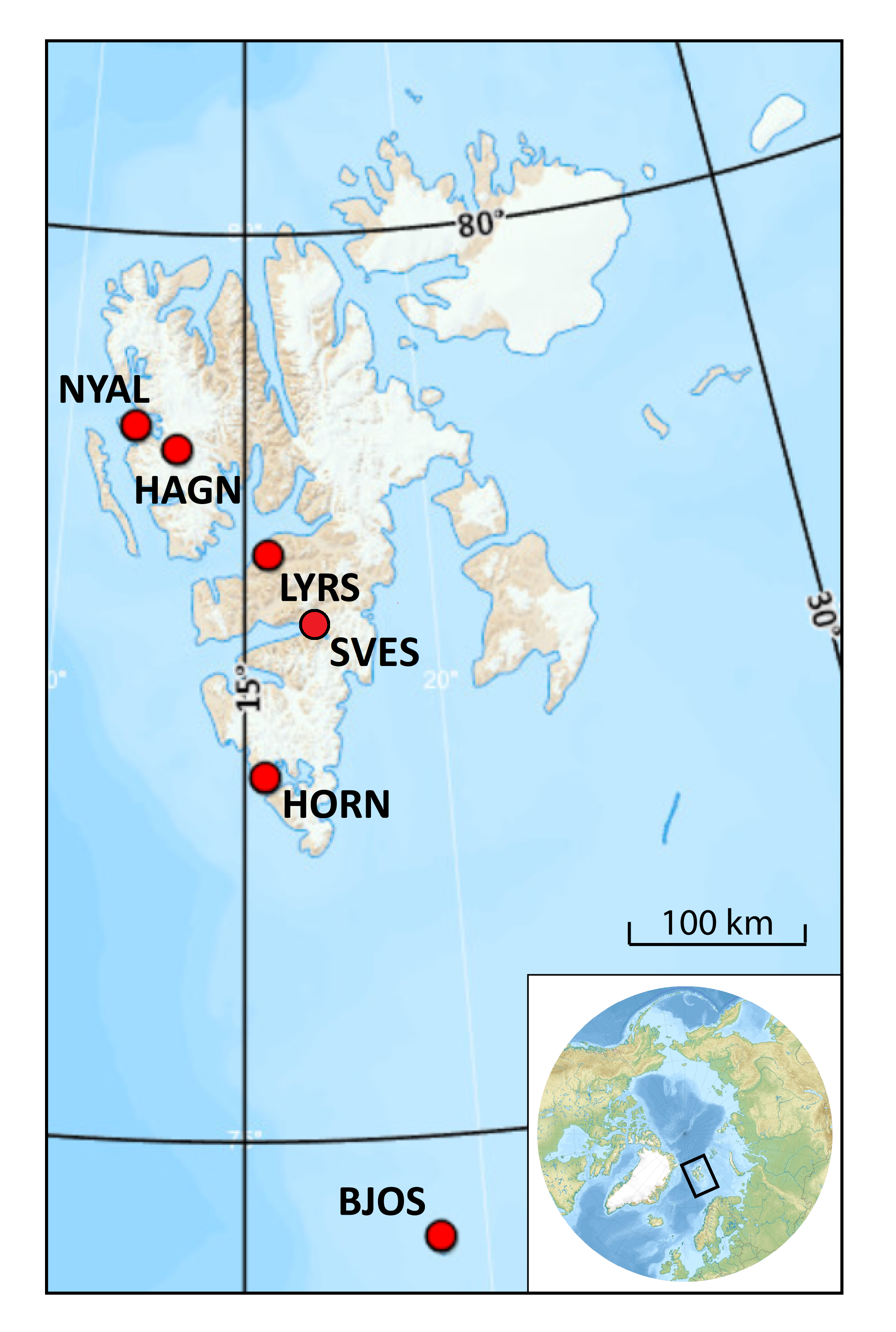}
\caption{Geodetic network on Svalbard. The location NYAL include the GNSS stations NYAL and NYA1, the VLBI antenna NYALES20 and the SCG instrument.}
\label{fig:network}
\end{figure}

The main questions in this paper are: (1) How well do GNSS and VLBI capture the seasonal signal from glaciers and snow in Svalbard? (2) Will refining the \LWS\ models with a \CMB\ model improve the loading predictions? To answer these two questions we have studied GNSS time series from six locations on Svalbard (see Fig.~\ref{fig:network}) and the VLBI antenna in Ny-\AA lesund. We have used different analysis strategies both for the GNSS and the VLBI data sets. We have also filtered our time series for \NTL\ and \CM\ signals to improve the regional accuracy. The model described in \cite{vanpelt++2019} simulates glacier \CMB\ and seasonal snow conditions, from which variations in loading from glaciers and snowpack are extracted. 

In Section~\ref{sec:data} we describe the different data sets used in this study. We describe the softwares and analysis strategies for geodetic analysis,  the time series analysis, the \CM\ filtering and the different models used for loading predictions. In Section~\ref{sec:results}, we compare the geodetic results with the loading signal from glaciers and snow, \ATM, \NTO\ and \LWS. Based on this we discuss possibilities and limitations in our solutions for revealing the seasonal elastic signal. We also study the effect of refining the hydrological model with the \CMB\ model (Subsection \ref{ssec:rishyd}).

\section{Data and data analysis}
\label{sec:data}

\subsection{CMB model}
\label{ssec:cmb}
Glacier mass change is primarily the result of surface - atmosphere interactions (affecting snow accumulation and melt), snow processes (affecting melt water retention and runoff) and frontal processes (calving and frontal ablation of tidewater glaciers). Glacier mass changes due to atmosphere - surface - snow interactions are described by the \CMB, which describes the mass change of a vertical column of snow/firn/ice, in response to surface mass, and energy exchange and runoff of melt water. The \CMB\ dominates seasonal glacier mass change, with mass gain from snow accumulation during the cold season and melt-driven mass loss during the melt season. 

\citet{noel++2020} have shown that for all glaciers in Svalbard the mass fluxes of precipitation (+23~Gt/yr) and runoff (-25~Gt/yr) dominate the seasonal climatic mass balance cycle in recent decades (1985-2018), with nearly all runoff concentrated in the summer months (June, July and August) and snow accumulating the rest of the year. These mass fluxes are much larger than the estimated mean ice discharge due to calving and frontal ablation from tidewater glaciers \citep[7~Gt/yr][]{Blaszczyk++2009}. Svalbard-wide constraints on the seasonality of combined calving and frontal ablation are currently lacking and not considered here. Previous estimates on three glaciers in Svalbard however indicate that frontal ablation is more substantial in summer and early autumn than during winter and spring \citep[][]{Luckman++2015}.

Here, we use the \CMB\ model data set, described in \cite{vanpelt++2019}, and extract weekly output for the period 1990-2018. \cite{vanpelt++2019} used a coupled energy balance - subsurface model \citep[][]{vanpelt++2012} to simulate \CMB\ for all glaciers in Svalbard, as well as seasonal snow conditions in non-glacier terrain. Both the glacier and seasonal snow mass changes are accounted for. They describe weekly mass changes resulting from snow accumulation, surface moisture exchange, melt and rain water refreezing, and retention in snow, and runoff.  Runoff estimates are local and no horizontal transport of water is accounted for. 

\subsection{Elastic loading signal}
\label{ssec:loading}
Mass redistribution results in Earth's crust deformation called mass 
loading \citep{r:darwin1882}.
Mass loadings are caused by the ocean water mass redistribution due to 
gravitational tides and pole tide (ocean tidal loading), by variations of 
the atmospheric mass (\ATM\ loading), by variations of the bottom
ocean pressure due to ocean circulation (\NTO\ loading), and
by variations of land water mass stored in soil, snow, and ice (\LWS\ loading). Mass loading crustal deformations have a typical magnitude
at a centimetre level \citep[see e.g.][]{petrov+boy2004}.

\citet{r:love1911} showed that the deformation caused by
mass loadings can be found in a form of an expansion into spherical
harmonics. Each spherical harmonic of the deformation field
is proportional to the spherical harmonic of the surface pressure 
exerted by loading mass. The proportionality dimensionless coefficients
called Love numbers that depend on a harmonic degree are found by 
solving differential equations.
Therefore, when the global pressure field 
mass redistribution is known, the elastic deformation can be found by expansion 
of that field into spherical harmonics, scaling the harmonics by Love numbers
and performing an inverse spherical harmonics expansion. 

Love numbers were computed using the REAR software \citep[][]{r:mel14}
  for the Earth reference model STW105 \citep[][]{kustowski++2008}. 
  Time series of \NTL\ from \ATM, \NTO\ and \LWS\ have been used in our analysis. Input to the \ATM\ loading is the pressure field from NASA's numerical weather model MERRA2 \citep[][]{galero++2017}.
  The \NTO\ loading uses the model MPIOM06 \citep[][]{jungclaus++2013}, 
and the \LWS\ loading uses the pressure field of MERRA2 model \citep[][]{r:merra_lws}. The MERRA2 model accounts for soil moisture at the depth of 0--2~meter 
and accumulated snow. 3D displacements cause by these loadings were computed using spherical harmonics transform of degree and order 2699 
and presented at a global grid $2'\times 2'$ with 
a time step of 3 or 6 hours. Then mass loading at a given
station is found by interpolation. The time series of these
loadings 
are available at the International Mass Loading Service \web{http://massloading.net}
\citep{r:imls15}.

However, the MERRA2 numerical weather model do not adequately describe
accumulation and runoff of water, snow, and ice at glaciers. It does
not consider all complexity of glacial mass change processes and its resolution, 16x55~km,
is insufficient to catch fine details in Svalbard. Here, we test the impact of replacing the above global model component for snow and ice with the regional snow and glacier \CMB\ product with  $1 \times 1$~km resolution. The model is described in Section~\ref{ssec:cmb}. We have re-gridded
the $1 \times 1$~km 
model to a uniform, regular, latitude-longitude grid with 
a resolution of $30''\times 30''$. The model value at a given element of the 
new grid is 

\begin{equation}
   M_{ij} = \frac{\strut \displaystyle\sum_{ab} M_{ab} e^{-r_{ij,ab}/D}} 
   {\displaystyle\sum_{ab} e^{-r_{ij,ab}/D}},
\end{equation}
where $M_{ab}$ is the model value for element {\it a,b}, the $r_{ij,ab}$ is the distance between grid points {\it i,j} and
{\it a,b}, and $D$ is the kernel distance set to 1~km.

We have computed mass loading time series, from 1990-08-05 through 2018-08-26 with a step of 7 days, at a  $30''\times 30''$ grid from the
\CMB\ output using spherical harmonic expansion degree and order
10799. This high resolution was used to correctly model the signal at stations that are located close to the edge of glaciers.

The choice of the degree/order of the expansion is determined by
availability of computing resources. The higher degree/order of the spherical
harmonic transform, the less errors near the coastal line. Atmospheric, 
land-water storage, and non-tidal ocean loading are computed with the time 
resolution of 3~hours and the total computation time using degree/order 
2699 is about 4 years per single core for CPUs produced in 2015--2020.
Since the glacier model has time resolution of 7 days we can afford 
to run computation with degree/order 10799 which allows to correctly 
model the signal at stations that are located close to the edge of glaciers.

  However, it is not sufficient to replace the \LWS\ loading 
computed on the basis of MERRA2 model with the mass loading computed on the 
basis of the \CMB\ model. Crustal deformation at a given point is affected by mass loading not only from the close vicinity, but also from remote areas. Therefore, in order to account for loading 
displacement caused by mass redistribution from the area beyond Svalbard
archipelago, we computed an additional series of \LWS\ loading using MERRA2 model that was set to zero outside Svalbard archipelago.
 The total \LWS\ loading
displacement is:
\begin{equation}
  \label{eq:merra_modi}
  D_{\rm LWS} = D_{\rm merra2} - D_{\rm merra2,svalbrard} + D_{CMB},
 \end{equation}
where $D_{\rm merra2}$ is the displacement from MERRA2 model,  $D_{\rm merra2,svalbrard}$ is the loading signal from the MERRA2 model that was set to zero  except latitude 
$76^\circ < \phi < 81^\circ$ and longitude $10^\circ < \lambda < 34^\circ$ (the area including the Svalbard archipelago) and $D_{CMB}$ is the displacement form the \CMB\ model. 
\begin{figure*}
   \includegraphics[width=0.48\textwidth]{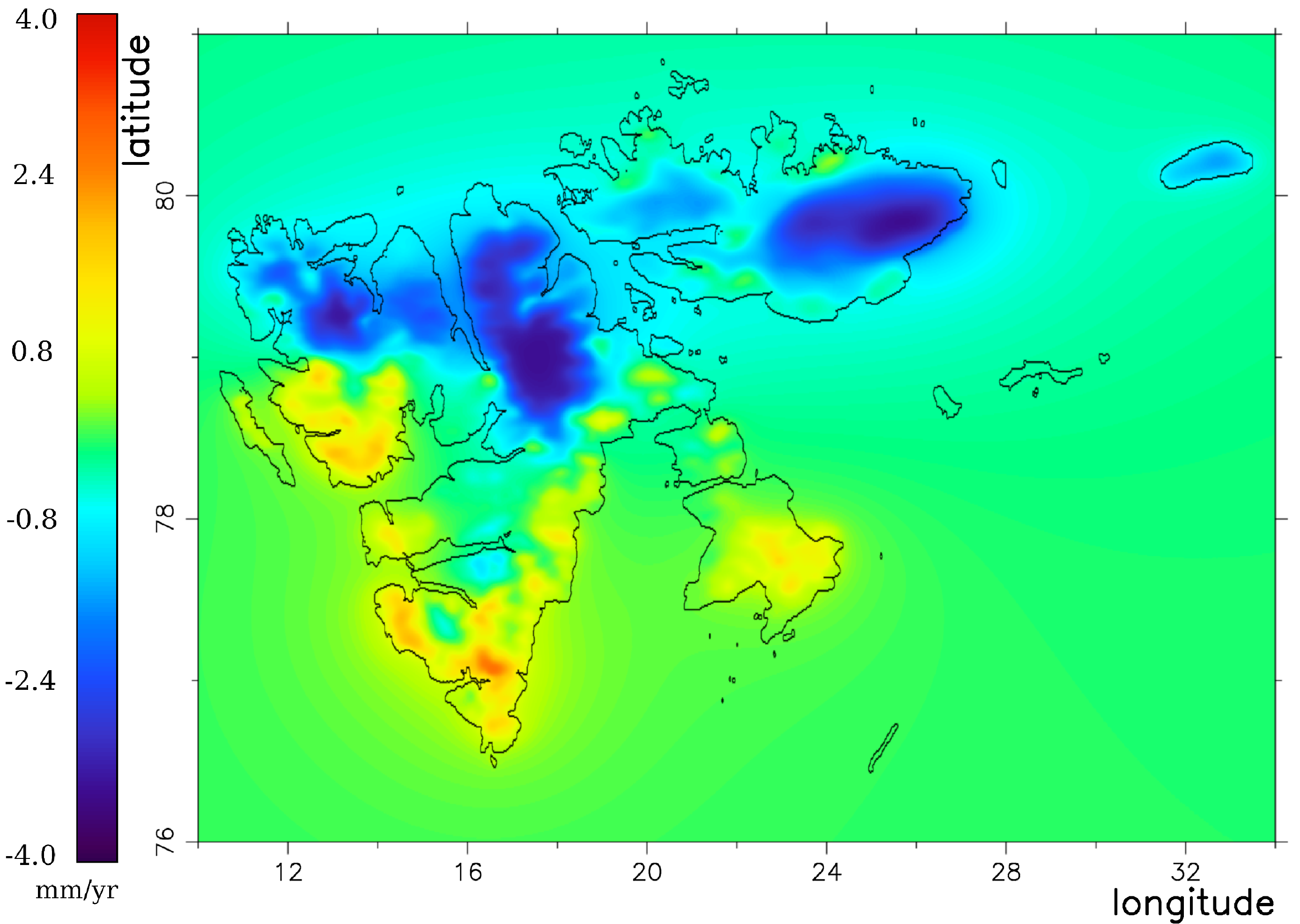}
   \hspace{0.0199\textwidth}
   \includegraphics[width=0.48\textwidth]{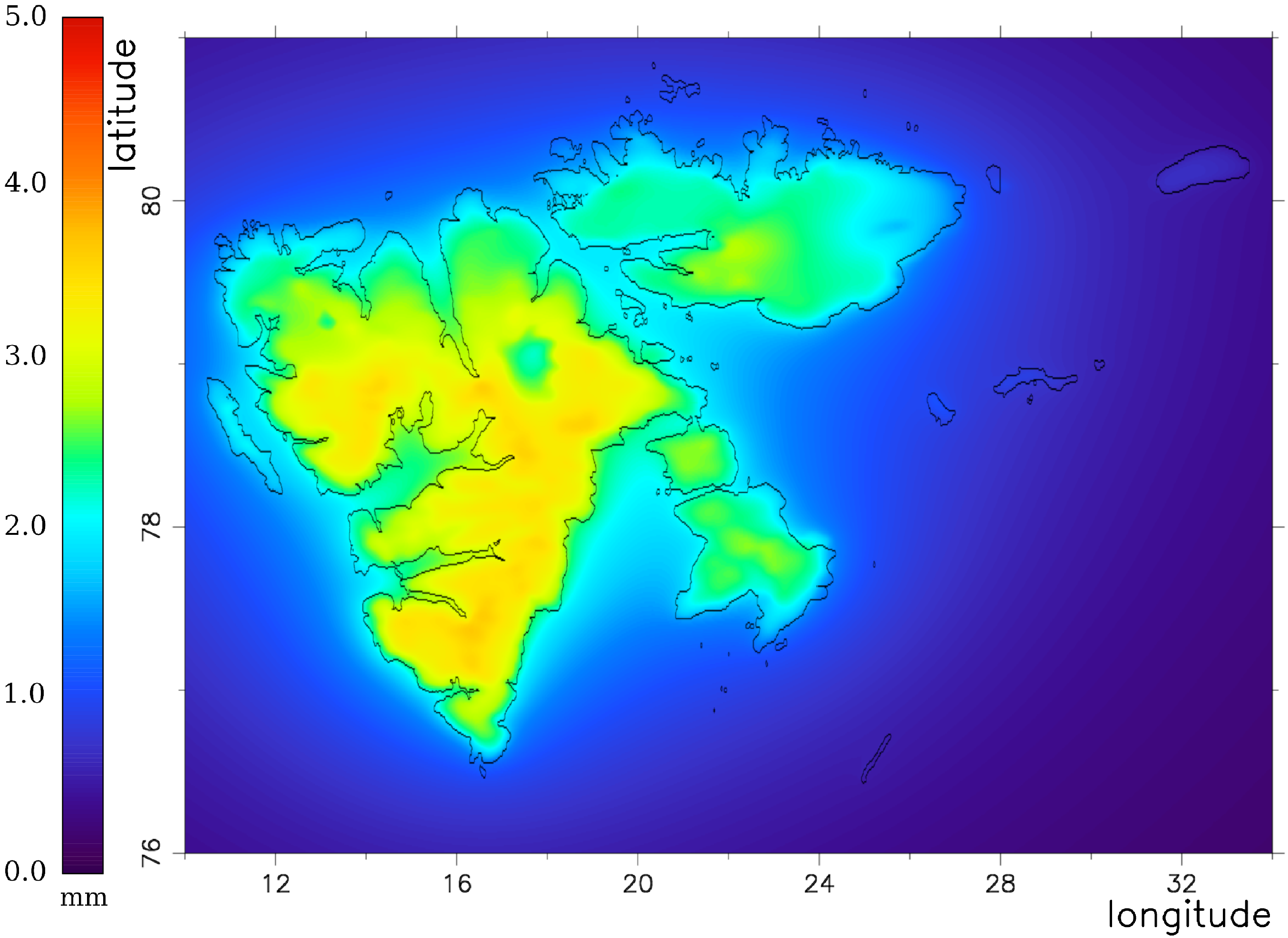}
\caption{Crustal deformations due to glacier and snow loading according to the CMB model. The panels are:  the rate of change (left) and annual signal (right). Important: the CMB model does not account for mass loss due to frontal ablation and calving. }
\label{fig:lws_loading}
\end{figure*}

Fig.~\ref{fig:lws_loading} shows the high-resolution maps of the rate and amplitude of the annual signal in crustal deformation caused by the water mass change in Svalbard archipelago according to the CMB model.
The parameters were estimated in a 4-parameter least square regression (mean value, rate and sine and cosine annual term) for the time series in each grid point.

\subsection{GNSS data analysis}
\label{sec:gps-analysis}
In this study we have used 30 seconds daily RINEX data re-sampled to five  minutes, from five permanent \GNSS\ stations on Svalbard (NYAL, NYA1, LYRS, SVES and HORN), and one station on Bear Island (BJOS) 240 km south of Svalbard (see Fig.~\ref{fig:network}). All stations are located close to existing settlements with infrastructure like power supply and communication. We have also used data from station, HAGN, located at a nunatak in the middle of the glacier Kongsvegen 30 km southeast of Ny-\AA lesund. This station is powered by solar panels and batteries. In the dark season, data is recorded for 24 hours once a week to save power until the sun is back. Data is downloaded during a field trip once a year.

\GNSS\ data are analyzed with the program packages Gamit/Globk \citep[][]{herring++2018} and GipsyX \citep[][]{bertiger++2020}. The GipsyX software is using undifferentiated observations. We are using the \PPP\ approach \citep[][]{zumberge++97} and the solutions are 
in the \IGS\ realization of \ITRF\ \citep[][]{altamimi++2014} through the \JPL\ orbit and clock products. We distinguish between the GispyX-FID and GipsyX-NNR solution, whereby either JPL fiducial (FID) or \NNR\ orbit and clock products are applied. The \NNR\ products are only constrained via three no-net-rotation parameters to the \ITRF\ solution, whereas the FID products are tied in addition with three translation and one scale parameter to \ITRF\ \citep[][]{bertiger++2020}. Gamit software uses double difference observations. To ensure a good global realization in \ITRF\ of the Gamit solution a global network of approximately 90 global \IGS\ stations was analyzed and combined with the Svalbard stations before transforming to \ITRF. The global stations were all stable stations with long time series. Daily coordinate time series are extracted from these solutions.

The two stations in Ny-\AA lesund belong to the \IGS\ network and are analyzed by several institutions, \UNR\ \citep[][]{blewitt++2018}, \JPL\ \citep[][]{heflin++2020}, and \SOPAC \citep[][]{bock+2012}. NYAL and NYA1 are also included in the latest \ITRF\ \citep[][]{altamimi++2014} solution. Key parameters for the different analysis strategies are given in Table~\ref{tab:strategies}.
  
\begin{table*}
  \caption{GNSS analysis strategies. (*) Elevation dependent site by site functions, where $a$ and $b$ are estimated based on postfit editing of residuals from each station. $E$ is the elevation angle.}
\label{tab:strategies}
\scriptsize{
\centering
\begin{tabular}{l|llllllll}\hline
&Gamit-NMA &	GipsyX-FID & GipsyX-NNR & Gamit-SOPAC&	GipsyX-UNR&	GipsyX-JPL\\ \hline
Orbit and clock product &Estimated& 			JPL fiducial &	JPL-NNR& Estimated&			JPL-NNR &	JPL-NNR\\
Elevation angle cutoff &10 degree&			7 degree&	7 degree  &10 degree&			7 degree&	7 degree \\
Elevation dependent weighting	& $a^2 + b^2/sin(E)^2$ (*) & $ 1/\sqrt{sin(E)}$ &	$ 1/\sqrt{sin(E)}$ 	& 	$ a^2 + b^2/sin(E)^2$ (*) &	$ 1/sin(E)$ &	$ 1/\sqrt{sin(E)}$\\
Troposphere mapping  function  & VMF1 & VMF1 &			VMF1  & VMF1 &			VMF1&	GPT2w\\
2nd order ionosphere model  & IONEX from CODE & IONEX from JPL &  IONEX from JPL & IONEX from IGS &	IONEX from JPL &  IONEX from JPL\\
Solid Earth tide &IERS2010 & IERS2010 & IERS2010&IERS2010 & IERS2010 & IERS2010\\
Ocean tidal loading &FES2004& FES2004 &			FES2004& FES2004 & FES2004 &	FES2004\\
Ocean pole tide &IERS2010  &IERS2010&	IERS2010   &IERS2010 & IERS2010  &	Not applied\\
Ambiguity& Resolved &Resolved&			Resolved& Resolved & Resolved& Resolved\\ \hline
\end{tabular}
}
\end{table*}

The time series are analyzed with Hector software \citep[][]{bos++2008}. We have used the following model function:
\begin{equation}
  \label{eq:trend}
  h(t)=A+B t + \sum_{j=1}^{2} C_j cos(j2\pi t-\phi_j),
\end{equation}
where $A$ is the constant term, $B$ is the rate, $C_j$ is the amplitudes of the sinusoidal constituents, and $\phi_j$ is the corresponding phases. 
We have assumed that the temporal correlation in the time series are a combination of white noise and flicker noise. We have used data from 2010-01-01 until 2018-10-01 in all the GNSS results and comparisons. This limited time period ensures that we have the same time period for all the stations (except HAGN which was established in 2013), no breaks due to equipment shift, and the time series overlap with the \CMB\ model (see Section~\ref{ssec:cmb}).

\begin{figure}
   \includegraphics[width=0.52\textwidth]{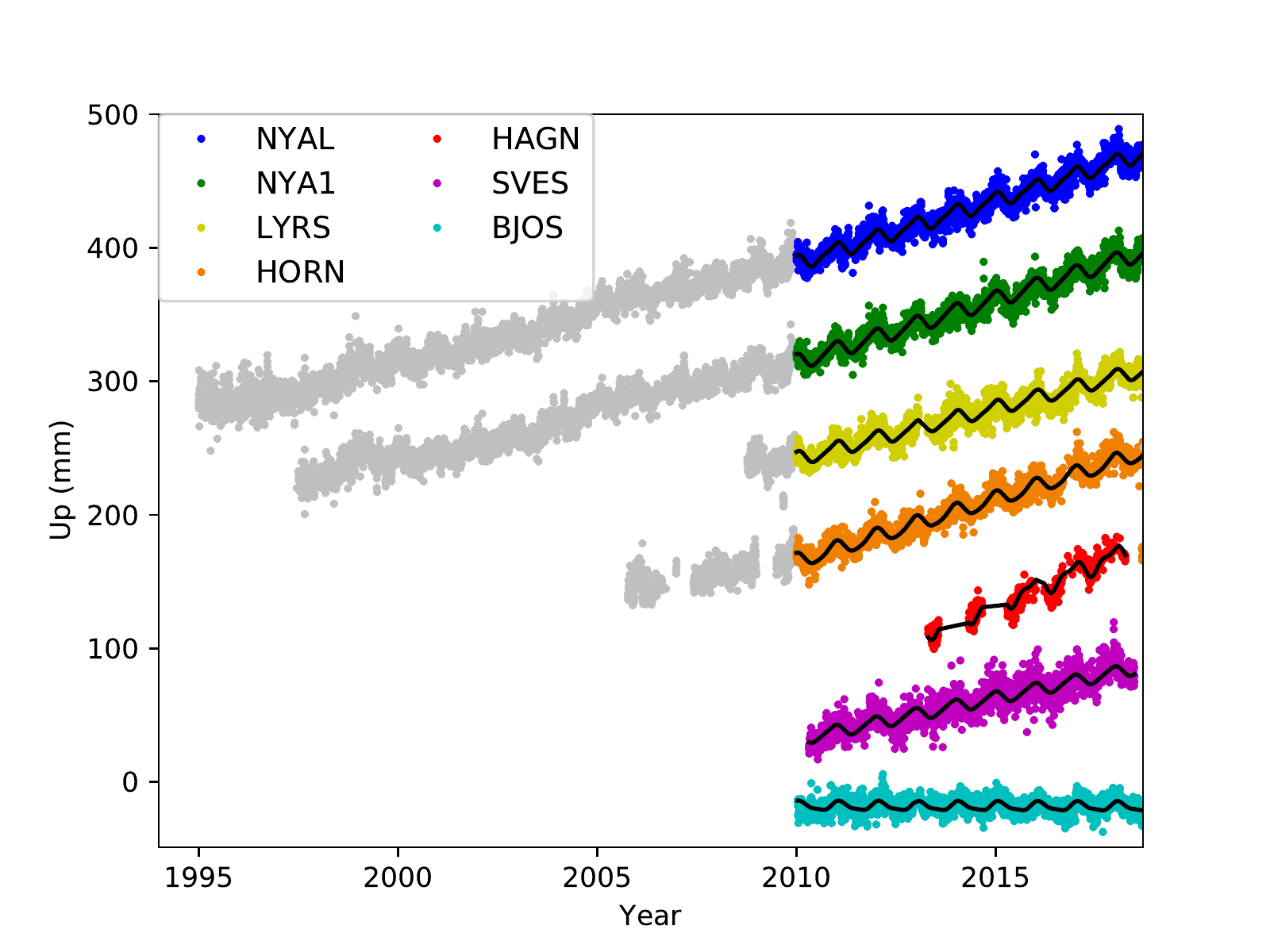}
\caption{GNSS vertical time series for the Gamit-NMA solutions. The color coded time span is the data period used in this study. The black curves are the model function fitted to this period. The HORN station was moved to a new location in 2009. The time series are shifted with respect to each other to improve readability.}
\label{fig:raw-ts}
\end{figure}

The time series for the vertical component of the Gamit-NMA solution is plotted in Fig~\ref{fig:raw-ts}.

\subsection{VLBI}
\label{ssec:vlbi}

 The VLBI station NYALES20 participated in 2183 twenty four hour observing sessions from 1994-10-04 to 2020-10-19. We ran several solutions.

Solution s1 was obtained using the geodetic analysis software Where 
\citep[see][for more details]{kirkvik++2017}. VLBI observing sessions were 
individually analyzed with the following approach:
a priori station coordinates were taken from \ITRF\ including the post-seismic 
deformation models or VTRF2019d (IVS update of ITRF2014) for newer stations. 
To define the origin and the orientation of the output station position
estimates, tight no-net-translation and no-net-rotation with respect to \ITRF\ were imposed. 
A priori radio source coordinates were taken from the ICRF3 S/X catalog \citep[][]{charlot++2020} and corrected 
for the galactic aberration. The source coordinates were not estimated. 
A priori Earth orientation parameters were taken from the C04 combined EOP 
series consistent with \ITRF. The Earth orientation parameters, polar motion, polar motion rate, UTC-UT1, length of day, and celestial pole offsets, were then estimated for each session. In addition, troposphere and clock parameters was estimated.Key parameters for the VLBI solutions are included in Table~\ref{tab:vlbi-strategies}.
\begin{table}
\caption{VLBI analysis strategies.}
\label{tab:vlbi-strategies}
\scriptsize{
\centering
\begin{tabular}{l|ll}\hline
  & Where & pSolve \\ \hline
  A priori radio source coordinates & ICRF3 S/X   & Solved for \\
  A priori EOP & C04 combined & Solved for                  \\
  Elevation angle cutoff & 0 degree & 5 degree                     \\
  Troposphere mapping function & VMF1 & Direct integration   \\
                       &  \citep[][]{boehm++2006}     & using output of   \\
                       &      & numerical weather    \\
                       &      & model GEOS-FPIT             \\
  Solid Earth tide  & IERS2010 & Elastic           \\
  & &  \citep{r:mdg97}           \\
  Ocean tidal loading & TPXO7.2 & FES2014B                        \\
  Ocean pole tide  & IERS2010 & IERS2010\\
  Higher order ionosphere & Not applied & Applied \\  \hline
\end{tabular}
}
\end{table}

Solution s2 was obtained using VLBI analysis software suite 
pSolve (\web{http://astrogeo.org/psolve}).
Source 
position, station positions, station velocity, sinusoidal position variations
at annual, semi-annual, diurnal, semi-diurnal frequencies of all the stations,
were estimated as global parameters in a single least square solution using all dual-band 
ionosphere-free combinations of VLBI group delays from 1980-04-12 to 
2020-12-07, in total 14.8 million observations. There are 28 stations that have 
discontinuities due to seismic events or station repair. These discontinuities and
associated non-linear motion was modeled with B-splines with multiple knots, and 
the B-spline coefficients were treated as global parameters.
In addition to global parameters, the Earth orientation parameters, pole coordinates,
UT1, their first time derivatives, as well as daily nutation offsets are 
estimated for each observation session individually. Atmospheric zenith path delay
and clock function are modeled with B-splines of the 1st degree with time span
60 and 20 minutes, respectively. A so-called minimum constraints on station positions
and velocities and source coordinates were imposed to invert the matrix of the
incomplete rank. These constraints require that the net translation and rotation
station positions and velocities of a subset of stations be the same as in ITRF2000
catalog and net rotation of the so-called 212 defining sources be the same 
as in ICRF. It should be noted that s2 solution is independent on the choice
of the a~priori reference frame, i.e. change in the a~priori position does not
affect results.

The data reduction model included modeling thermal variation of all antennas, 
oceanic tidal, \NTO\, \ATM\,  and \LWS\ loading with one exception, where for 
station NYALES20 the following \LWS\ model were used $D_{\rm merra} - D_{\rm merra,svalbard}$. Implying that the a priori
model totally ignores mass loading exerted by water mass redistribution
in Svalbard.

The VLBI network is small and heterogeneous: different stations participate
in different experiments. Therefore, the time series of station position should
be treated with a great caution: the estimate of the position change of station 
X affects the position estimate of station Y because of the use of the net 
translation and  net rotation constraints to solve the system of the incomplete
rank. An alternative approach to processing time series is estimation of 
admittance factor. We assume that the time series of the displacement 
in question d(t)  is present in data as $a \cdot d(t)$ where $a$ is a
dimensionless parameter called an admittance factor that is assumed constant 
for the time period of observations. The admittance factor describes what
share of the modeled signal is present in observations.

\begin{figure}
   \includegraphics[width=0.490\textwidth]{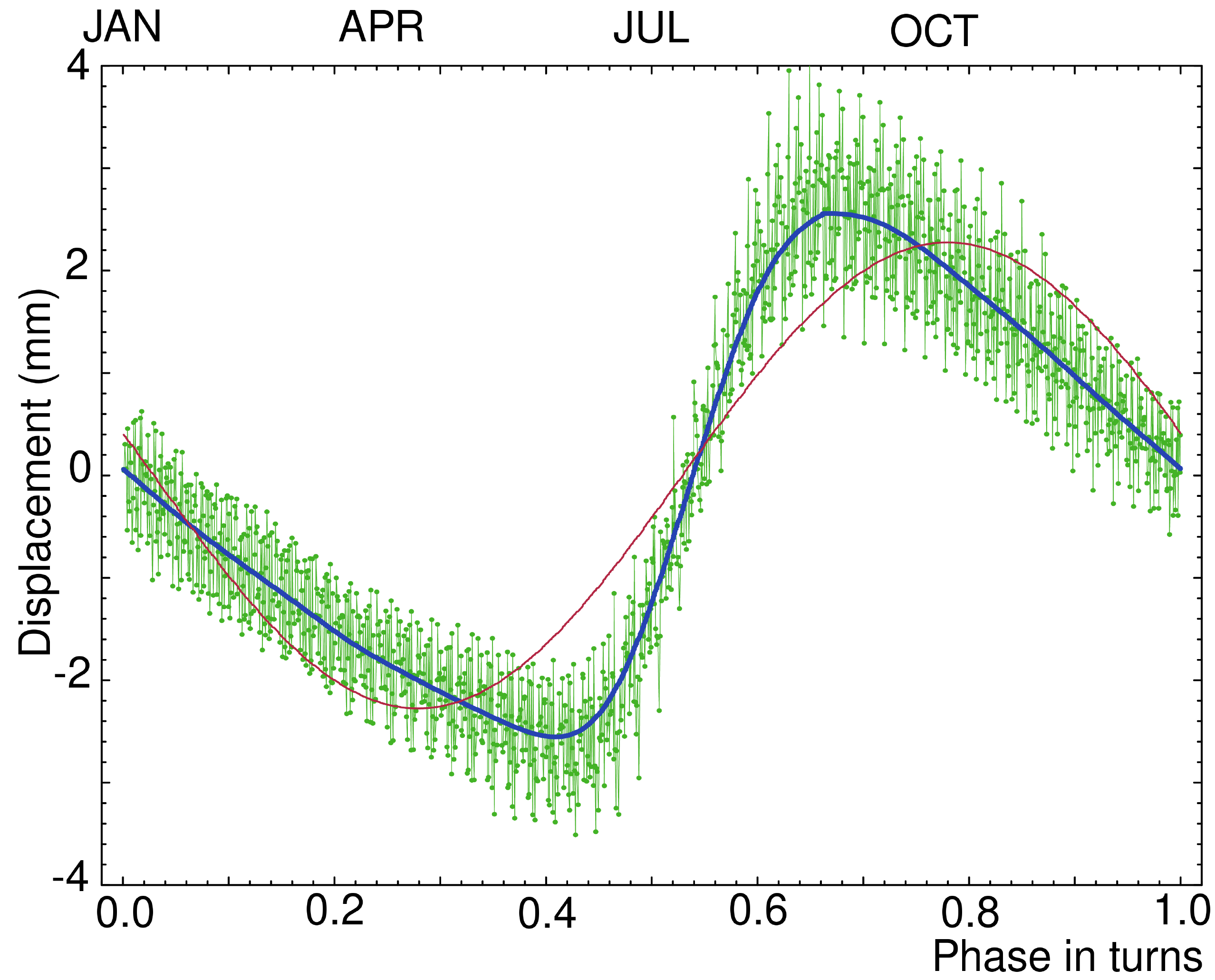}
   \caption{Folded periodic up LWS mass loading
            displacements of NYALES20 after removal of the slowly varying 
            constituent. The thick blue line shows the estimate of
            the seasonal constituent. Green dots show the mass loading signal
            after removal of the interannual constituent. A red thin line
            shows a sinusoidal fit in a form 
            $a \cos 2\pi \, p + b \sin 2\pi \,p$, where $p$ is the
            phase of the seasonal signal in turns.
        }
\label{fig:lws_nyales20_sea}
\end{figure}

We noticed that seasonal crustal deformations of NYALES20 positions are 
periodic but not sinusoidal. The shape of these 
variations is surprisingly stable with time (See Fig.~\ref{fig:lws_nyales20_sea}). We decomposed the 
mass loading signal into four components: seasonal, interannual, 
linear trend, and residuals. The decomposition was performed in 
three steps. First, the mass loading time series were filtered 
with the low-pass Gaussian filter, which provided a coarse interannual 
signal (IAV$(t)$). Second, the time series  were folded of the phase in a form 
$p=(t - t_0)/\Delta t$, where $t$ is time, $t_0$ is the reference epoch
2000.0, $\Delta t$ is the period (one year), and then smoothed. That 
provided a coarse estimate of the seasonal signal (SEA$(t)$, blue 
curve in Fig.~\ref{fig:lws_nyales20_sea}).
Then we adjusted parameters $A$, $B$, $a_i$, $s_i$ of the 
decomposition of the loading displacements $D(t)$ described by the 
Eq~\ref{e:dec}, using a single least square solution:

\begin{equation}
    D_{\rm CMB}(t) = {\rm IAV}(t) + {\rm SEA}(p(t)) + A + Bt + \varepsilon(t),
  \label{e:dec}
\end{equation}
where
\begin{eqnarray*}
    {\rm IAV(t)}  & = & \sum_i a_i B_i(t)     \\
    {\rm SEA(t)}  & = & \sum_i s_i B_i(p(t)),
\end{eqnarray*}
where $B_i$ is the basis spline of the 3rd degree with the pivotal 
knot $i$.

\begin{figure}
   \includegraphics[width=0.50\textwidth]{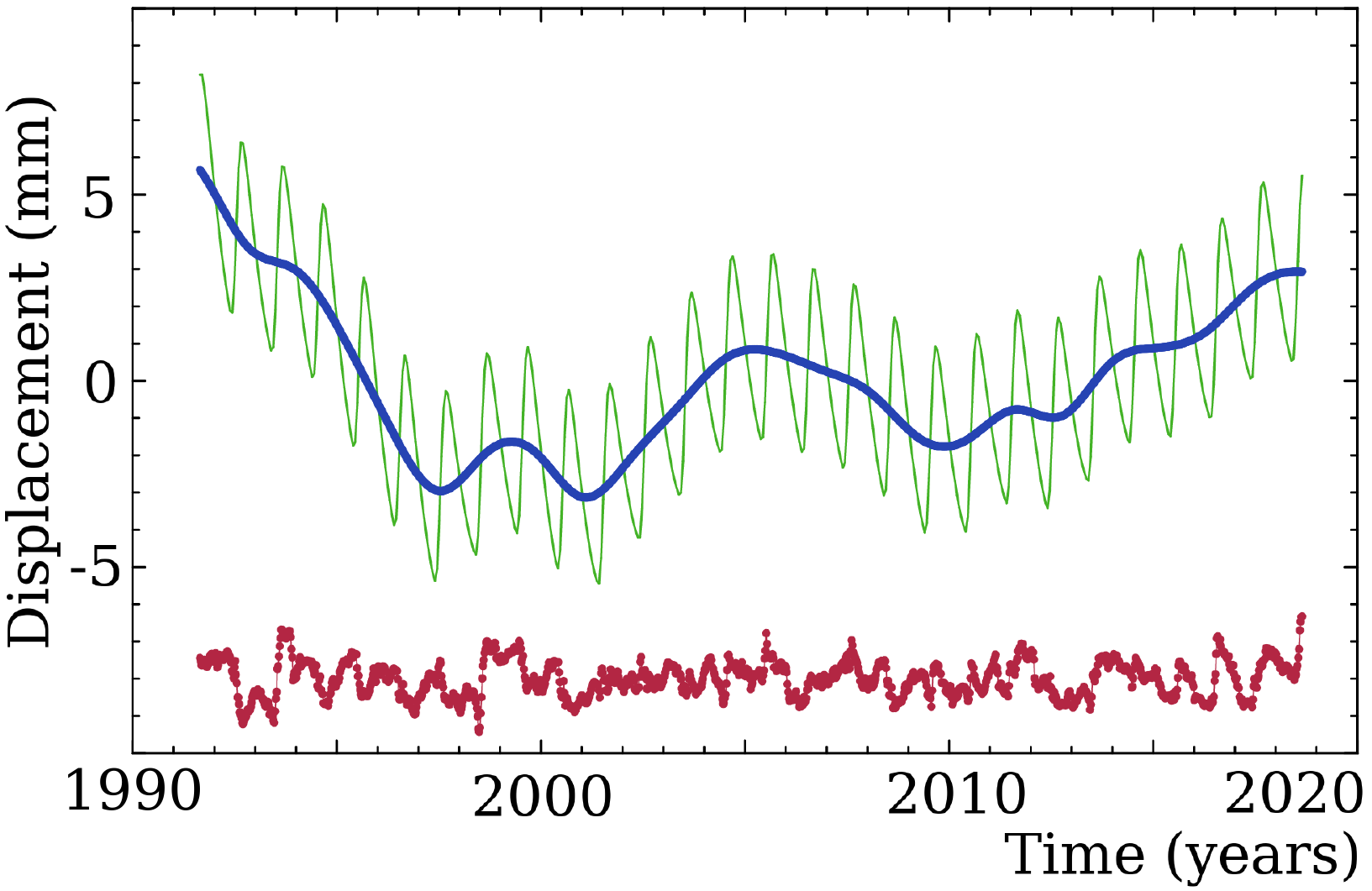}
   \caption{Three constituents of the vertical LWS
            mass loading at station NYALES20. The thick blue line shows
            the interannual variation, the green thin line shows the 
            seasonal component, and red dots in the bottom shows 
            the residual signal. The residual signal is artificially
            shifted by -8mm. The linear trend is removed and not shown.
        }
\label{fig:lws_loading_up}
\end{figure}

Fig.~\ref{fig:lws_nyales20_sea} illustrates the seasonal 
component of the loading signal at NYALES20. A thin red line at the 
plot shows result of the best fit of the sinusoidal signal. 
However, the sinusoidal model provides a poor fit to the data with errors 
reaching 40\% of the seasonal signal. All constituents of this 
expansion for NYALES20 are shown in Fig.~\ref{fig:lws_loading_up}.

In solution s3 we did not estimated annual and semi-annual 
sinusoidal variations of NYALES20 positions, but estimated 
admittance factors for the up, east, and north components of the 
${\rm IAV}(t) + {\rm SEA}(p(t))$ mass loading time series. 
In contrast to estimation of sinusoidal variations, the shape and phase of the 
signal remains fixed when we estimate admittance. The adjusted
parameter is the scaling factor of the modeled displacement 
magnitude. The power of this approach is that it allows us to evaluate
quantitatively the amount of the modeled signal in data,
and test a statistical hypothesis that all model signal 
is present in the data.

  The results of admittance factor estimation are presented 
in Table~\ref{tab:adm} in row ADM\_TOT. Then we estimated the 
admittance factor for the seasonal ${\rm SEA}(p(t))$ and interannual 
variations ${\rm IAV}(t)$ separately in the s4 solution.

\begin{table}
   \caption{Admittance factors of NYALES20 displacements caused by 
            LWS loading. The first row, ADM\_TOT
            shows the admittance factor estimate from s3 solution of the 
            total mass loading signal. Rows ADM\_SEA and ADM\_IAV shows 
            estimates of the seasonal and interannual constituents of the 
            loading signal from s4 solution respectively.
           }
   \begin{center}
      \begin{tabular}{l r r r}
          \hline
          Factor     & \multicolumn{1}{c}{Up} &  \multicolumn{1}{c}{East} &  \multicolumn{1}{c}{North} \\ \hline
          ADM\_TOT   &  1.38 $\pm$ 0.04  &  0.62 $\pm$ 0.05 &   2.05 $\pm$ 0.12 \\
          ADM\_SEA   &  1.10 $\pm$ 0.05  &  0.47 $\pm$ 0.11 &   6.10 $\pm$ 0.49 \\
          ADM\_IAV   &  2.90 $\pm$ 0.07  &  2.44 $\pm$ 0.11 &   1.00 $\pm$ 0.15 \\
          \hline
      \end{tabular}
   \end{center}
   \label{tab:adm}
\end{table}

\subsection{Gravimetry/SCG}
\label{ssec:gravimetry}
We use gravity measurements from two \SCG\ instruments covering the period 1999 to 2018
to estimate gravity change. Gravity measurements from 1999 to 2013 and 2014 to 2018 are
collected with C039 and iGrav012 SCG instrument, respectively. The original gravity
measurements have a spacing of 1 second, giving a total of approximately 620 million measurements.
They are re-sampled every minute using a symmetric numerical Finite Impulse Response (FIR) zero
phase low-pass filter with a cut-off at 120 seconds \citep{Wenzel1996}. Data was then cleaned for
outliers and earthquakes. We corrected for the effect of air pressure using the value of -0.422 $\pm$ 0.004 $\mu$Gal/hPa found by \cite{sato++06b}. Both the solid earth and ocean tides are removed from the gravity data by estimating a synthetic tide based on \cite{Hartmann+Wenzel1995} tidal model and a set of tidal parameters. The synthetic tide is estimated using ETERNA 3.4 \citep{Wenzel1996}.

We also estimated and removed the instrumental drift by comparing to absolute gravity measurements. We estimated a linear drift (using unweighted least squares) by comparing to ten AG measurements (2000, 2001, 2002, 2004, 2007, 2010, twice in 2012, 2014, 2017). The estimated value is -2.74107 $\pm$ 0.17 $\mu$Gal/year.
Finally, we re-sampled the data first every 5 minutes and then every 1 hour using a symmetric FIR zero phase filter (cutoffs 1250 sec and 2 hours, respectively) and then to daily values using a flat filter.

\subsection{Filtering of Common Mode and elastic loading signal}
\label{ssec:ECM}
It is well known that stations in a region can have a spatially correlated signal, a so-called \CM\ signal \citep[][]{wdowinski++97}, and that removal of the \CM\ signal can reduce noise in the time series. The \CM\ signal could come from the \GNSS\ analysis strategy and from the strategy for reference frame realization. It could come from mismodeled orbit, clocks or EOPs, or through unmodeled large scale hydrology or atmospheric effects. To remove such signal either \CM\ filtering, \EOF\ or regional reference frame realization, can be used. All these methods presuppose that we have stations exposed to the same undesirable \CM\ signal. In Arctic areas, we have limited access to nearby stations. All stations on Svalbard are exposed to similar signals from glaciers, using one or several of these stations for removal of the \CM\ signal will not only remove the \CM\ signal, but also the real elastic signal from snow and ice.

The station BJOS at Bear Island is located 240 km south of Svalbard. The Island is small and surrounded by ocean and the local loading signal from ice and snow is approximately 10\% of the signal in Ny-\AA lesund (see Tabel~\ref{tab:sval-loading-detailed}). It is the closest GNSS station outside Svalbard. Time series of the BJOS station are used to estimate the \CM\ signal. Time series for the BJOS station, and hence \CM\ filtering, are only included in the solutions computed by the authors (Gamit-NMA, GipsyX-FID and GipsyX-NNR) and not available in the external solutions (SOPAC, JPL, UNR and ITRF).The \CM\ filtered time series for the $i$-th station is then:
\begin{equation}
H_{CM}^i(t)= H_{GNSS}^i(t)-CM(t) = H_{GNSS}^i(t)-H_{GNSS}^{BJOS}(t),
\end{equation}
where $t$ is the epoch and $H_{GNSS}^{i}$ and  $H_{GNSS}^{BJOS}$  is the time series for station $i$ and $BJOS$ respectively.

The \CM\ filtering removes the common error signal at the stations as well as real measured signal at BJOS. If the station at Bear Island has an unique unmodelled loading signal not present in other Svalbard stations, this unique signal will be erroneously subtracted also from the other stations.

To \CM\ filter a time series where the signal from a loading model is removed, the loading signal for the station(s) used in the \CM\ filtering has to be removed as well. In our case, the loading signal was subtracted both for the BJOS time series before computing the \CM\ signal and for the other Svalbard time series before the \CM\ filtering. The final Svalbard time series are cleaned for both the regional \CM\ signal over Svalbard and Bear Island and the estimated load signal. The \CM\ filtered time series for station $i$ is then:
\begin{equation}
  \label{eq:cm}
H_{CM,L}^i(t)= H_{GNSS}^i(t)-H_{L}^i(t)-CM(t),
 \end{equation}
where $t$ is the epoch, $H_{GNSS}^i$ is the observed time series, $H_{L}^i$ is the estimated loading signal, and $CM$ is the common mode signal. As described earlier, we use the time series from BJOS to estimate the \CM\ signal, but since we remove the estimated loading signal from the time series, we have to remove the loading signal from BJOS time series before computing the \CM\ signal. Therefore, we get
\begin{equation}
  \label{eq:cm-time-series}
  H_{CM,L}^i(t)=H_{GNSS}^i(t)-H_{L}^i(t)-(H_{GNSS}^{BJOS}(t)-H_{L}^{BJOS}(t)).
\end{equation}

\subsection{Isolating the elastic signal from glacier and snow}

The signal in $H_{CM,L}(t)$ (Eq.~\ref{eq:cm-time-series}) includes all vertical motions not accounted for in the loading models or \CM\ filtering, e.g., unmodeled loading, \GIA, tectonics, and noise. Assuming that the \GIA\ and the tectonic component are linear, the left hand side can be written $H_{CM,L}(t)=LIN(t)+\varepsilon(t)$, where $LIN$ is the linear part and $\varepsilon$ contains the noise. The noise includes unmodeled loadings, but also station dependent effects like multipath, atmospheric effects, not use of individual antenna calibration, and thermal expansion of antenna monument. Possible unique unmodelled signal from BJOS will also map into the noise term.  Splitting the load signal into a signal from glacier and snow, $H_{GS}$, and other non tidal loadings, $H_{NTL^*}$, we can rewrite Eq.~\ref{eq:cm-time-series} into: 

\begin{eqnarray}
  \label{eq:ris}
  LIN^i(t)+H_{GS}^i(t) + \varepsilon(t) & = & H_{GNSS}^i(t)-H_{NTL^*}^i(t) \nonumber\\
  & - & \left( H_{GNSS}^{BJOS}(t)-H_{NTL^*}^{BJOS}(t) \right. \nonumber \\
  &  & - \left. H_{GS}^{BJOS}(t) \right) , 
\end{eqnarray}
i.e. we have isolated the linear part and the elastic signal from glaciers and snow as a sum of known terms.

\section{Results and Discussion}
\label{sec:results}

The vertical component of the different GNSS solutions in Ny-\AA lesund 
and Bear Island as well as the \NTL\ signal are included in 
Table~\ref{tab:nyal-up}. In addition,  the Where results (Solution s1) from 
the NYALES20 VLBI antenna and the \SCG\ in Ny-\AA lesund are included. 
Some of the time series are plotted in Fig.~\ref{fig:time-series}. 
The horizontal components of the different GNSS solutions are included in Appendix, Table~\ref{tab:nyal-horizontal}.
The loading signals for all the different loading models are 
included in the Appendix, Table~\ref{tab:sval-loading-detailed}.
 
\begin{figure*}
\centering
 \includegraphics[width=.45\textwidth]{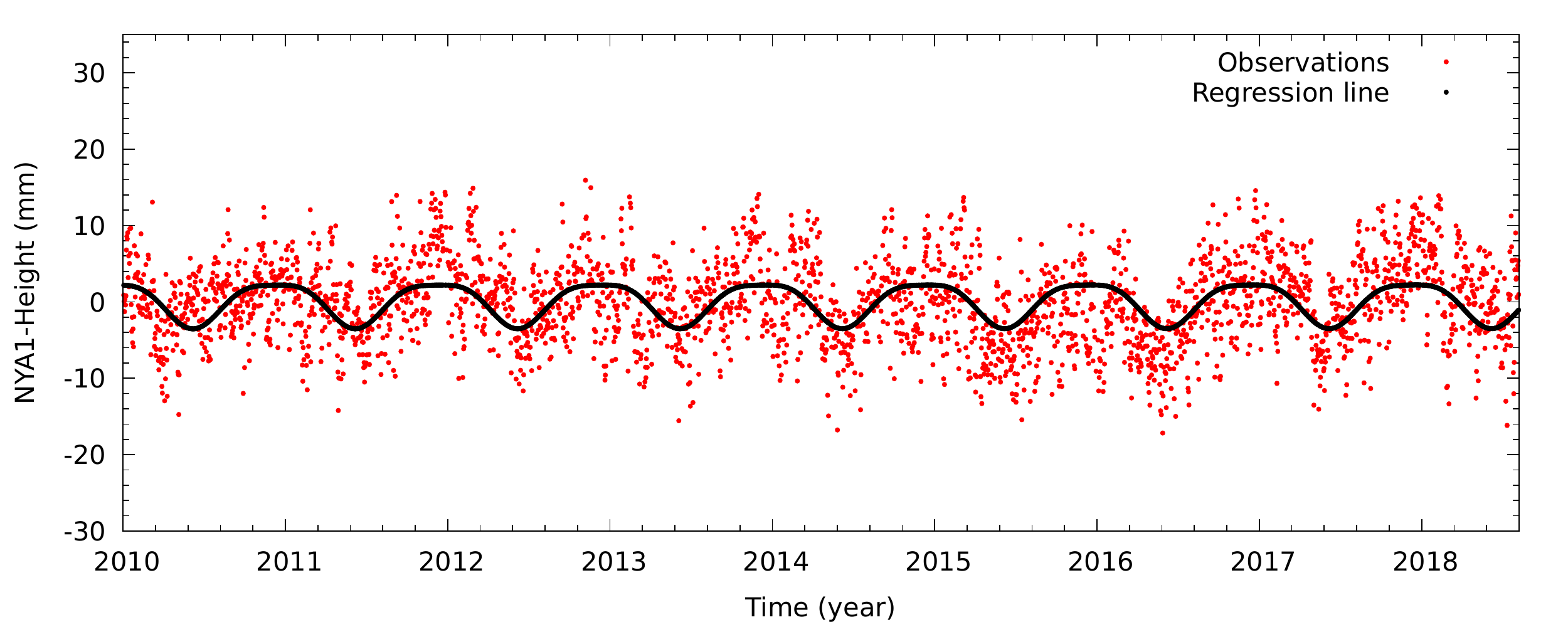}
 \includegraphics[width=.45\textwidth]{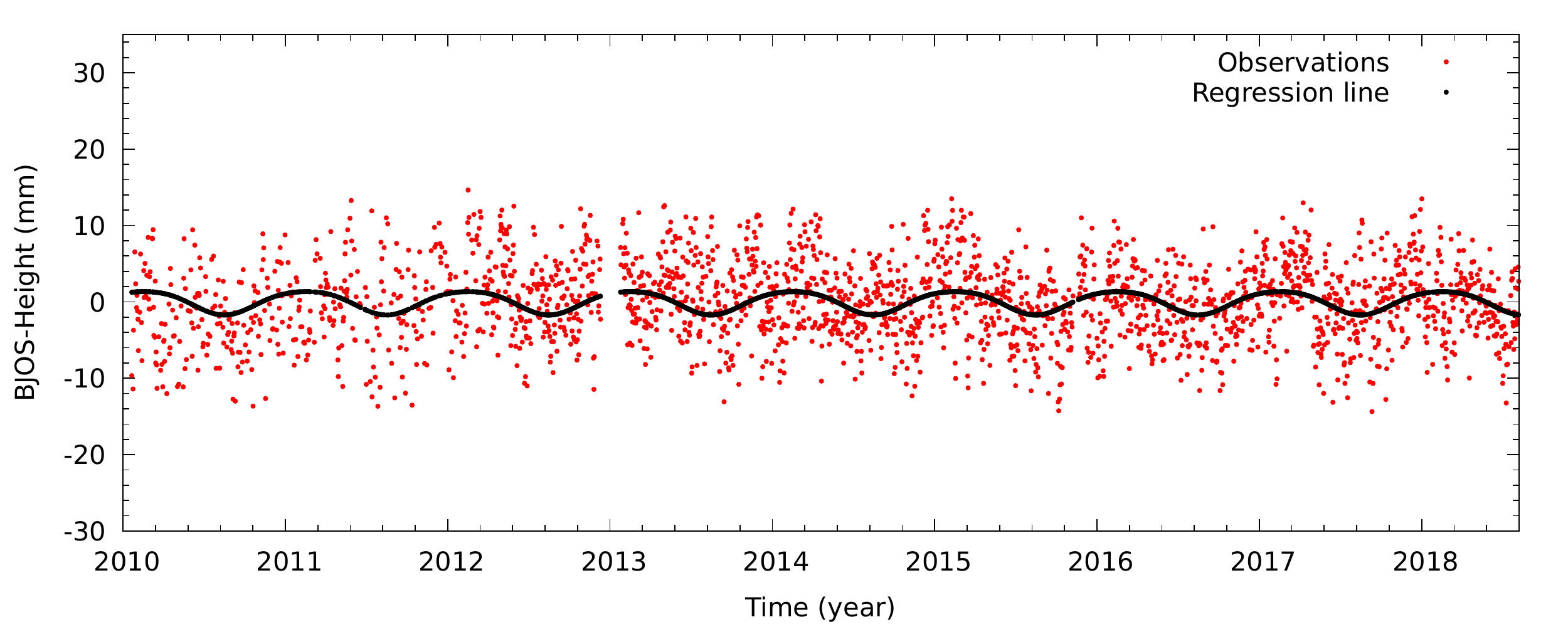}
 \includegraphics[width=.45\textwidth]{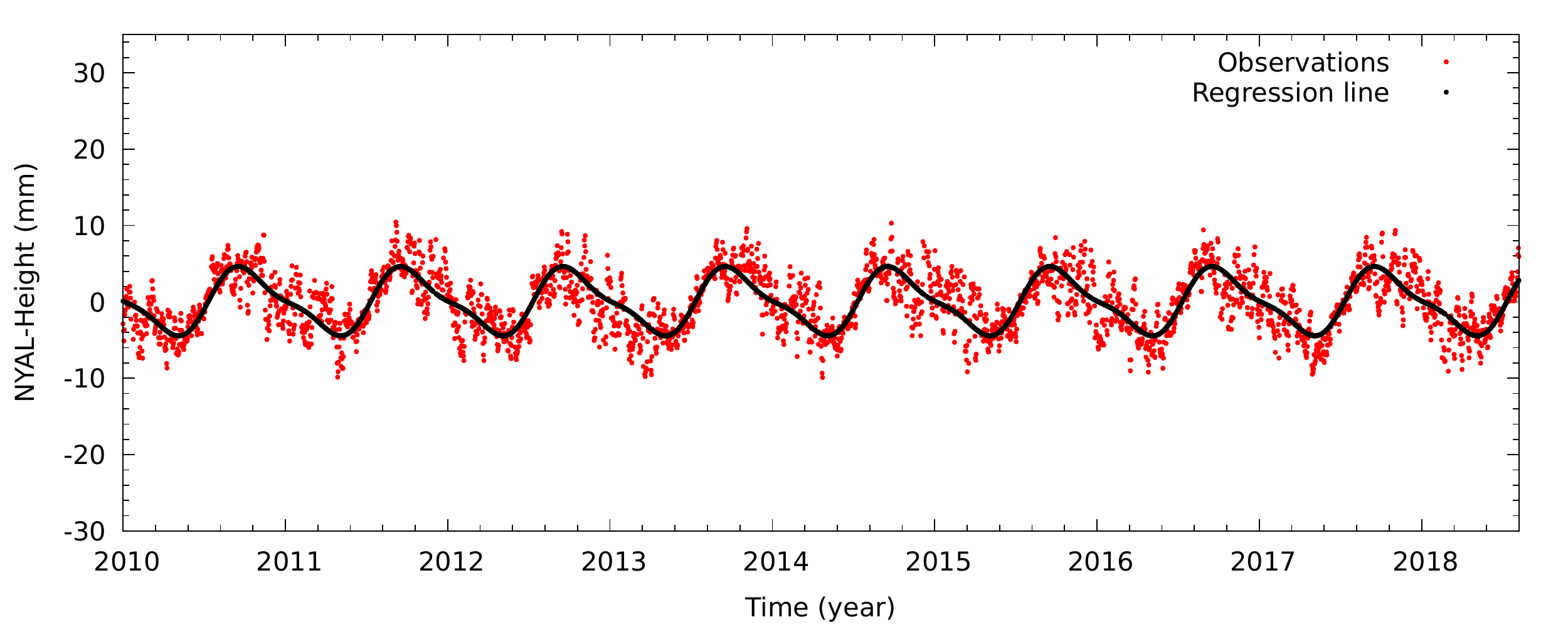}
 \includegraphics[width=.45\textwidth]{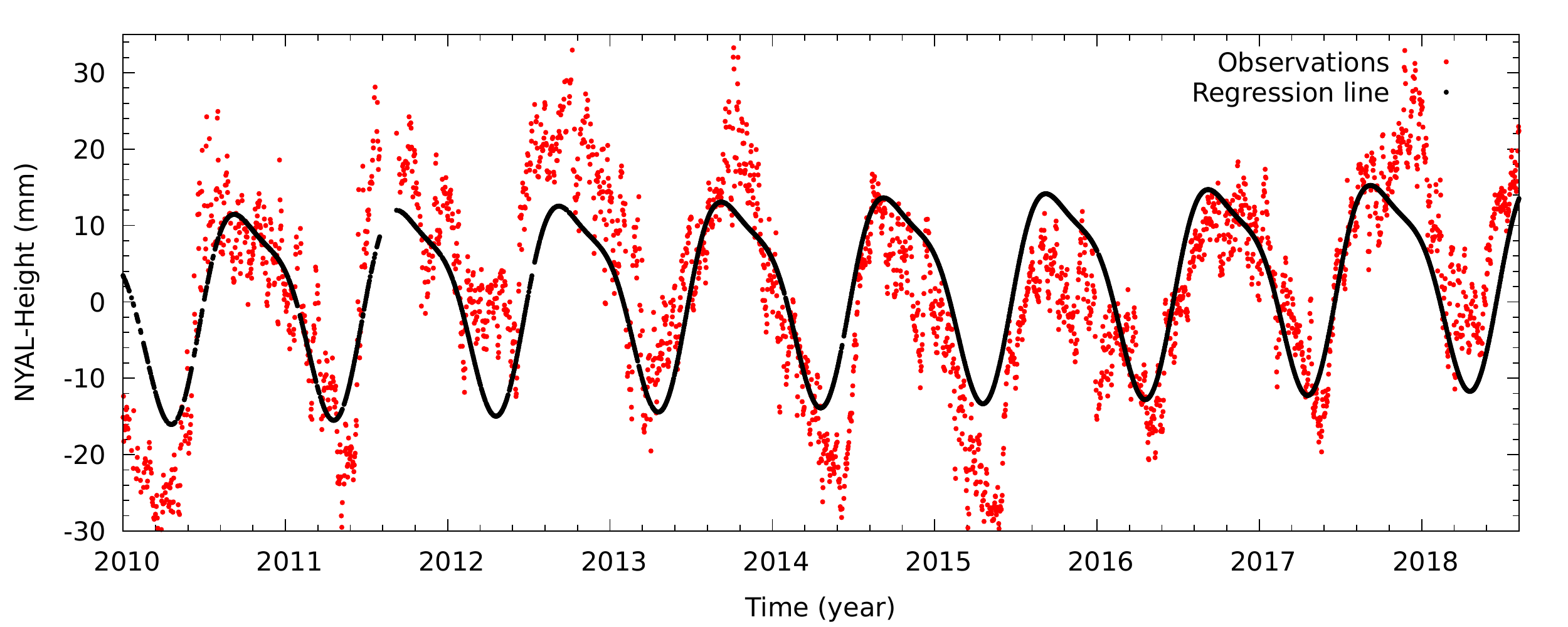}
 \includegraphics[width=.45\textwidth]{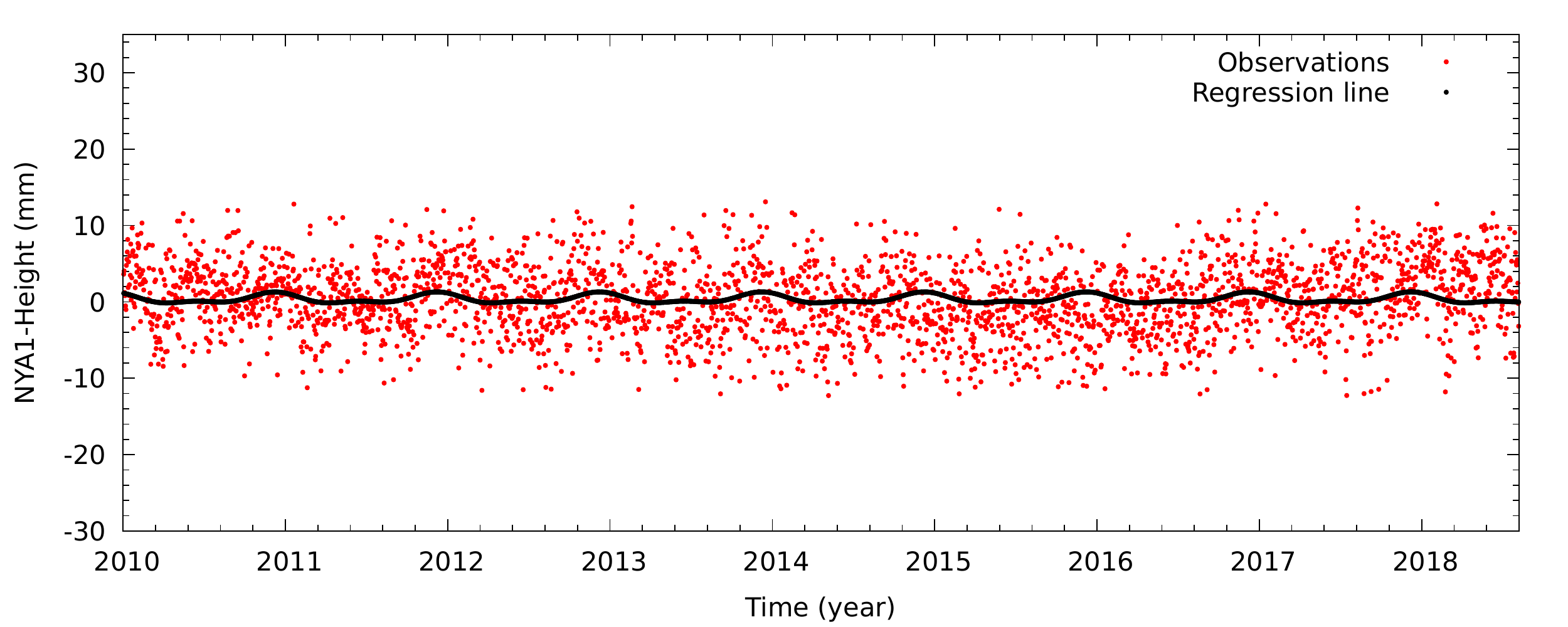}
 \includegraphics[width=.45\textwidth]{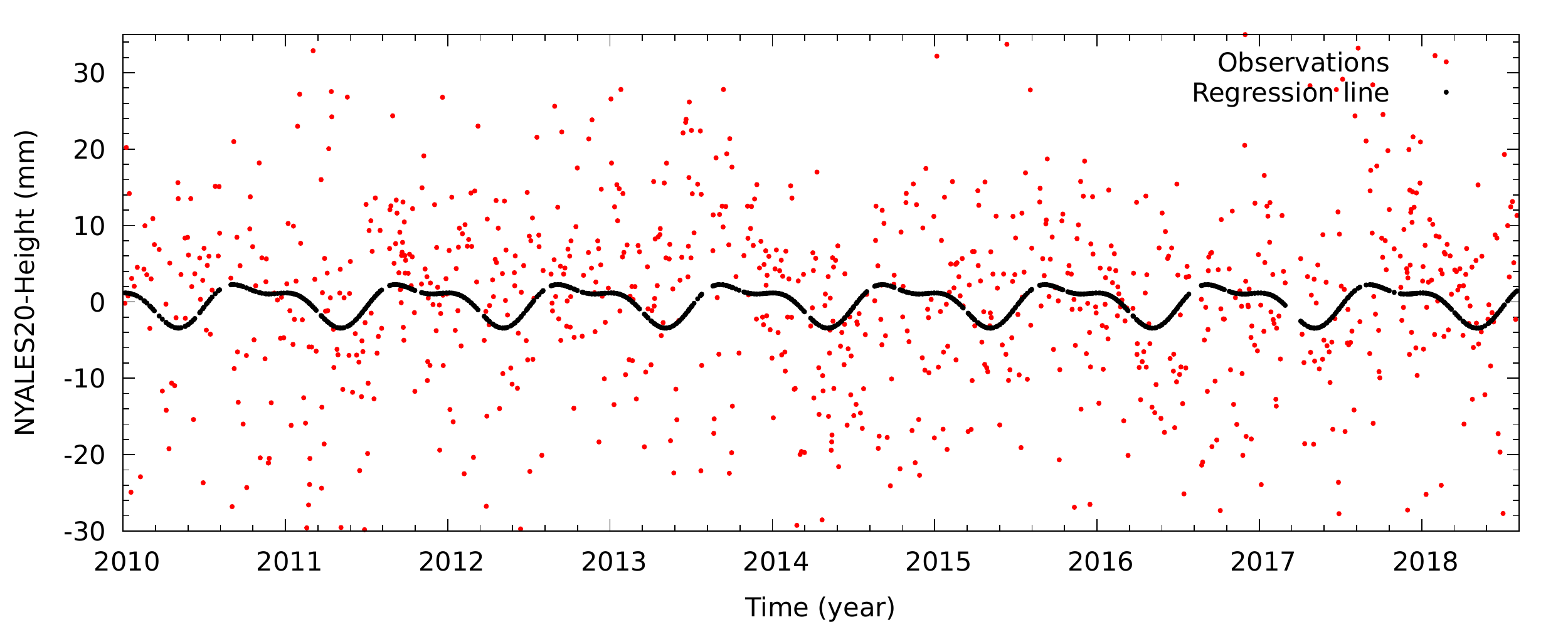}
 \caption{A selected set of detrended time series for Svalbard. The time series are: GipsyX-NNR for NYA1 (upper left), \NTL\ (\ATM, \NTO, \LWS\ including the glaciers and snow) signal in Ny-\AA lesund (middle left), the GipsyX-NNR time series for NYA1 after removal of \NTL\ and CM filtering (lower left), GipsyX-NNR for BJOS (upper right), gravity from the \SCG\ in Ny-\AA lesund (middle right) and the NYALES20 Where solution (lower right). The gravity values are converted to millimeter using the convertion ratio -0.24~$~\mu$Gal/mm \citep[][]{memin++2012}.}
\label{fig:time-series}
\end{figure*}

\begin{table*}
\caption{Trend, annual- and semi- signal in Ny-\AA lesund and Bear Island. The parameters 
  are estimated trend and annual signal estimated using Eq.~\ref{eq:trend}). The results are for different GNSS solutions, VLBI, SCG and NTL in Ny-\AA lesund and Bear Island. In the VLBI time series a pure white noise model is assumed.  The gravity values ($*$) are converted to millimeter using the convertion ratio -0.24~$~\mu$Gal/mm from \cite{memin++2012}. CM is the CM filtered time series described in Section~\ref{ssec:ECM}. NTL is the sum of non tidal elastic loading signal from ATM, NTO and LWS including the load from snow and glacier from the CMB model.}
\label{tab:nyal-up}
\footnotesize{
\centering
\begin{tabular}{llrrrrrrr}\hline
  Station & & Trend   & \multicolumn {2}{c}{Annual signal}& \multicolumn {2}{c}{Semi-annual signal}            \\ 
        & & (mm/yr) & Amp. (mm) & Pha. (deg) & Amp. (mm) & Pha. (deg)   \\ 
  \hline
NYA1 & Gamit-SOPAC &  9.61 +/-  0.62 &   6.28 +/-   0.64 &  -51.3 +/-     5.8 &   1.71 +/-   0.44 &  107.1 +/-   14.6   \\ 
     &  Gamit-NMA &  9.62 +/-  0.62 &   5.80 +/-   0.64 &  -13.0 +/-     6.3 &   1.42 +/-   0.44 &   57.6 +/-   17.2   \\ 
     & GipsyX-FID &  9.49 +/-  0.69 &   3.05 +/-   0.70 &  -45.7 +/-    13.0 &   1.05 +/-   0.45 &  123.1 +/-   23.2   \\ 
     & GipsyX-NNR &  9.26 +/-  0.67 &   2.96 +/-   0.69 &  -27.4 +/-    13.1 &   0.91 +/-   0.42 &  134.0 +/-   24.8   \\ 
     & GipsyX-UNR &  9.27 +/-  0.67 &   2.91 +/-   0.69 &  -27.6 +/-    13.3 &   0.91 +/-   0.42 &  135.2 +/-   24.8   \\ 
     & GipsyX-JPL &  9.59 +/-  0.65 &   3.36 +/-   0.66 &  -12.9 +/-    11.1 &   1.08 +/-   0.43 &  161.5 +/-   21.9   \\ 
     &   ITRF2014 &  9.00 +/-  0.95 &   4.05 +/-   0.75 &  -36.2 +/-    10.5 &   1.08 +/-   0.47 &  157.5 +/-   23.6   \\ 
     &  Gamit-NMA (CM) &  9.55 +/-  0.36 &   4.07 +/-   0.38 &  -47.5 +/-     5.4 &   0.66 +/-   0.26 &  113.1 +/-   21.3   \\ 
     & GipsyX-FID (CM) &  9.80 +/-  0.31 &   2.84 +/-   0.34 &  -54.0 +/-     6.8 &   0.86 +/-   0.24 &  136.5 +/-   15.9   \\ 
     & GipsyX-NNR (CM) &  9.86 +/-  0.35 &   2.91 +/-   0.38 &  -58.2 +/-     7.4 &   0.82 +/-   0.27 &  127.9 +/-   18.1   \\ 
\hline 
NYAL & Gamit-SOPAC &  9.41 +/-  0.61 &   6.24 +/-   0.63 &  -56.6 +/-     5.8 &   1.96 +/-   0.44 &  105.8 +/-   12.7   \\ 
     &  Gamit-NMA &  9.57 +/-  0.67 &   5.22 +/-   0.69 &  -12.9 +/-     7.6 &   1.75 +/-   0.48 &   64.3 +/-   15.4   \\ 
     & GipsyX-FID &  9.34 +/-  0.67 &   3.45 +/-   0.74 &  -59.4 +/-    12.1 &   1.16 +/-   0.48 &  122.1 +/-   22.5   \\ 
     & GipsyX-NNR &  9.14 +/-  0.66 &   3.19 +/-   0.68 &  -39.9 +/-    11.9 &   1.12 +/-   0.45 &  124.2 +/-   21.6   \\ 
     & GipsyX-UNR &  9.13 +/-  0.65 &   3.17 +/-   0.66 &  -39.9 +/-    11.8 &   1.11 +/-   0.44 &  125.5 +/-   21.6   \\ 
     & GipsyX-JPL &  9.39 +/-  0.65 &   3.44 +/-   0.66 &  -27.7 +/-    10.9 &   1.17 +/-   0.44 &  153.2 +/-   20.8   \\ 
     &   ITRF2014 &  9.34 +/-  0.98 &   4.37 +/-   0.78 &  -47.7 +/-    10.1 &   1.17 +/-   0.50 &  156.1 +/-   23.0   \\ 
     &  Gamit-NMA (CM) &  9.52 +/-  0.35 &   3.60 +/-   0.37 &  -52.9 +/-     5.9 &   0.96 +/-   0.26 &  102.8 +/-   15.3   \\ 
     & GipsyX-FID (CM) &  9.67 +/-  0.33 &   3.34 +/-   0.38 &  -63.9 +/-     6.5 &   1.17 +/-   0.28 &  122.9 +/-   13.4   \\ 
     & GipsyX-NNR (CM) &  9.75 +/-  0.34 &   3.45 +/-   0.36 &  -68.1 +/-     6.0 &   1.11 +/-   0.26 &  120.6 +/-   13.4   \\ 
\hline 
NYALES20 &    Where &  8.87 +/- 0.17 &   2.62 +/-  0.80 &  -67.3 +/- 17.9 &   1.14 +/-  0.81 &  77.3 +/- 31.5 \\ 
NYAL-SCG &       & $^*$2.52 +/- 0.64 & $^*$14.38 +/- 0.67  &  -83.8 +/- 2.7 &  $^*$3.92 +/- 0.47 & 60.6 +/- 6.8 \\ 
Ny-\AA lesund & NTL &  0.92 +/-  0.30 &   4.00 +/-   0.32 &  -82.5 +/- 4.6 & 1.21 +/- 0.22 & 111.3 +/-10.3  \\ 
\hline 
BJOS &  Gamit-NMA &  0.10 +/-  0.54 &   3.30 +/-   0.55 &   32.6 +/-     9.5 &   1.14 +/-   0.38 &   33.6 +/-   18.4   \\ 
     & GipsyX-FID & -0.27 +/-  0.62 &   0.90 +/-   0.47 &   21.2 +/-    27.3 &   0.62 +/-   0.32 &   45.5 +/-   27.5   \\ 
     & GipsyX-NNR & -0.47 +/-  0.59 &   1.63 +/-   0.58 &   46.1 +/-    19.6 &   0.57 +/-   0.30 &  -54.0 +/-   27.6   \\ 
\hline 
Bear Island   & NTL & -0.04 +/-  0.31 &   1.99 +/-   0.32 & -107.7 +/- 9.1 & 0.38 +/- 0.18 & 100.1 +/-25.3 \\ 
\hline 
\end{tabular}
}
\end{table*}

\begin{figure*}
\centering
 \includegraphics[width=.45\textwidth]{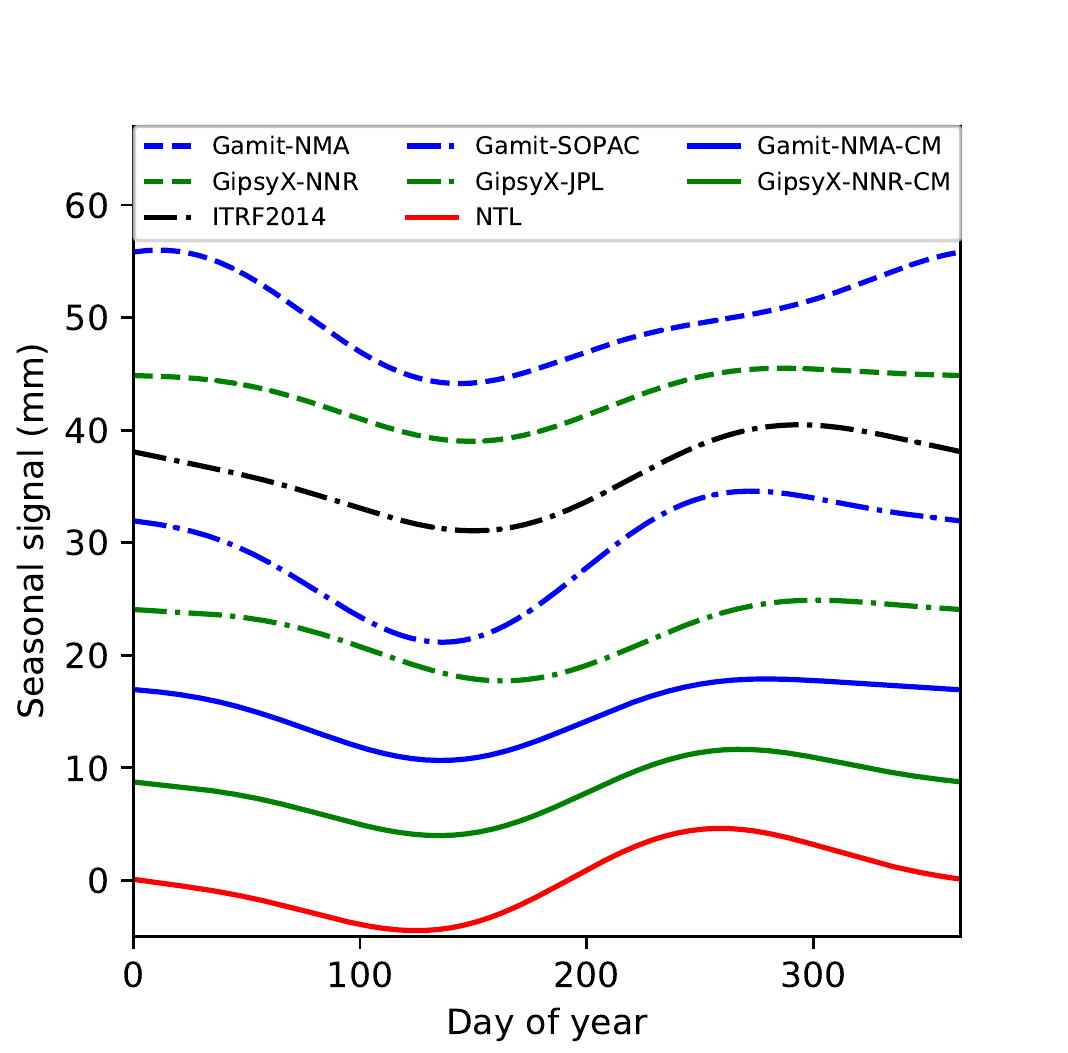}
 \includegraphics[width=.45\textwidth]{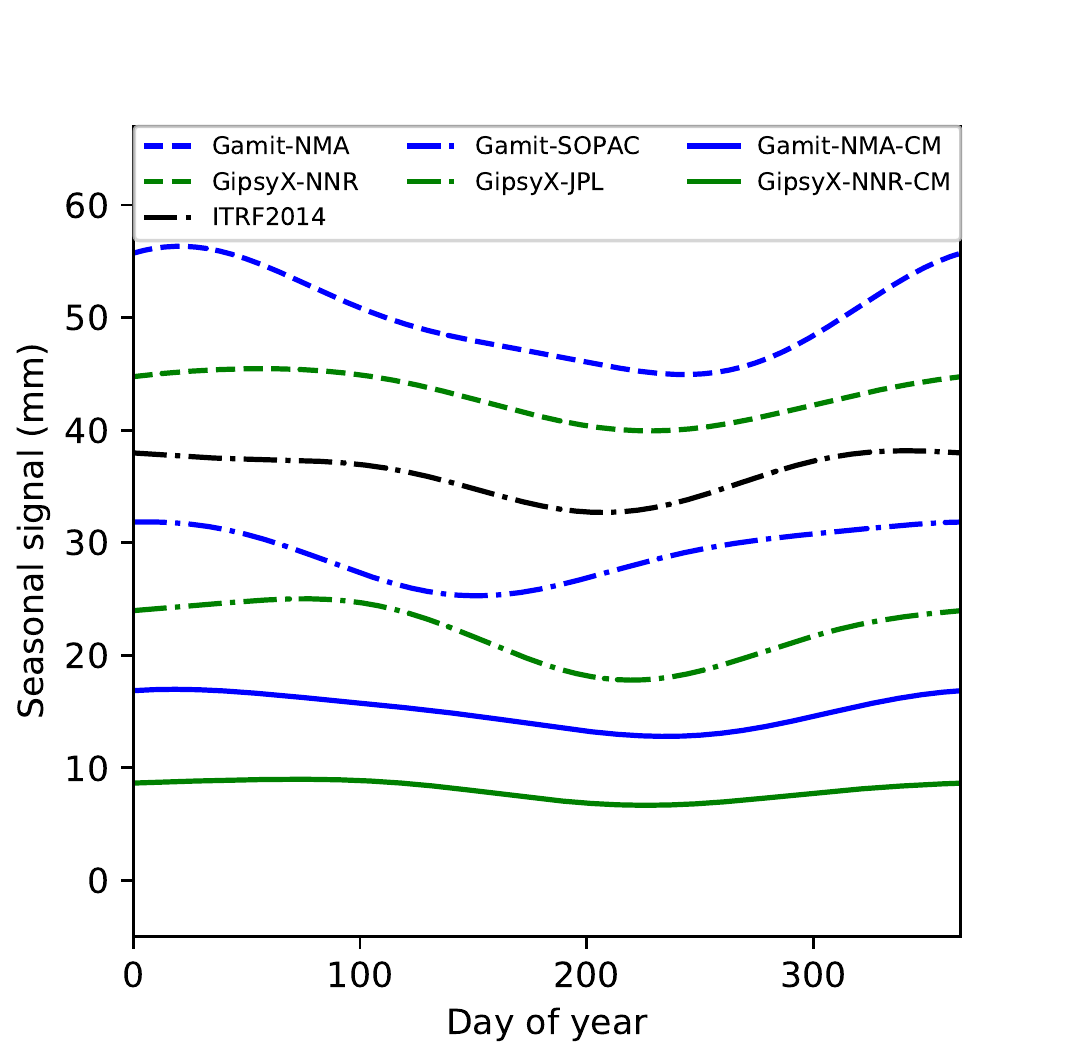}
  \caption{Seasonal signal in Ny-\AA lesund (NYA1). The left panel shows the sum of the annual and semi-annual sinusoidal signal for the time series, the right panel shows the same results relative to the NTL signal (ATM, NTO, LWS including the glaciers and snow). The upper most five curves are from time series analysis of the raw time series. The sixth and seventh curves are CM-filtered time series. The bottom curve of the left panel is the estimated NTL signal. The curves are shifted with respect to each other to improve readability.}
\label{fig:periodic-signal}
\end{figure*} 

\subsection{Determination of the loading annual signal}

As shown in \cite{kierulf++2009a} the uplift in Ny-\AA lesund varies from year to year. Consequently, trends from different time periods can not be compared directly. We have chosen to use the time interval 2010 until 2018 for time series analysis, except for the \ITRF\ time series that ended in 2014. The estimated uplift for the Ny-\AA lesund stations agree below the uncertainty level.

The annual signal in Ny-\AA lesund varies between the solutions both in phase and amplitude (Table~\ref{tab:nyal-up}, Table~\ref{tab:nyal-horizontal} and Fig.~\ref{fig:periodic-signal}). This implies that the choice of GNSS analysis strategy has a noticeable impact on the estimated seasonal variations. \cite{martens++2020} found similar differences in the estimated annual signal when they compared GNSS time series, in US and Alaska, based on different analysis strategies. Such variations make direct geophysical interpretation of the periodicity in GNSS time series difficult.

The measured vertical annual signal (Table~\ref{tab:nyal-up}) is smaller than the estimated \NTL\ signal for the GipsyX solutions and larger than the estimated \NTL\ signal for the Gamit solutions. The phase of the GNSS solutions are delayed relative to the \NTL\ signal with between 30 and 70 degrees (corresponding to a delay between one and two and a half months) in Ny-\AA lesund.

The \CM-filtered solutions are closer to the expected signal from \NTL\ and we have less differences between the GipsyX and Gamit solutions, see Fig.~\ref{fig:periodic-signal}. The annual signal found with the Where software for VLBI has a smaller amplitude, but the phase is close to the phase estimated from the loading modeling.

The phase of the gravity signal in Ny-\AA lesund is close to the phase of the loading models. The annual amplitude in gravity is $3.45~\mu Gal$. For a spherical and compressible Earth model the elastic gravity variations can be converted to vertical position variations using a ratio of -0.24~$~\mu$Gal/mm \citep[][]{memin++2012}. Using this ratio we got an yearly amplitude of 14.4~mm. This is much larger than the estimated annual loading signal of around 4.0~mm in Ny-\AA lesund.  However, the gravity variations also depends on the direct gravitational attraction. \cite{memin++2012} discussed how the location of the load affect the ratio between gravity and uplift. Both the distance to and the relative height of the load have an impact. The large annual signal implies mass changes at locations with negative relative heights close to the station.

The \SCG-instrument in Ny-\AA lesund gives a combined signal from three glacier related factors. The visco-elastic response from past ice mass changes, the immediate elastic response of the ongoing ice mass changes, and the direct gravitational attraction from the ongoing ice mass changes on the glaciers \citep[see][]{memin++2014,breili++2017}. The two latter have a clear influence on the annual signal. In addition, soil moisture and accumulated snow close to and mainly below the gravimeter, have a much stronger effect on gravity than on displacements.

Quantifying the gravity signal from these nearby hydrological factors are demanding and out of the scope of this paper. However, they, as well as glaciers, are forced by temperature and  precipitation. We assume that they are in phase with the elastic uplift signal. A gravimeter measures gravity changes directly, while VLBI and GNSS evaluate site positions from analysis of observations at a network, and a position estimate of a given station in general depends on 
measurements at other stations of the network. The phase of the \SCG\ time series is therefore an independent measure of the variations in Ny-\AA lesund and the result coincides with the results from the other techniques.

\subsection{Determination of the loading admittance factors}

We found the admittance factor from VLBI solutions for the 
seasonal vertical displacement does not deviate from 1.0 at a $2 \sigma$ level, 
i.e. the \LWS\ signal is fully  recovered from the data. 
At the same time the departure of the admittance factor from 1.0 for 
the horizontal loading components implies there is a statically 
significant discrepancy between the computed loading signal
and the data. It should be noted that the magnitude of the seasonal 
signal in North direction is only 0.15~mm and the signal itself
is just too small to be detected. The admittance factor for the 
interannual signal is significantly different from 1.0, which 
indicates that the loading signal alone cannot explain it.

We made an additional analysis to find the admittance factor for the glacier and snow loading signal at the GNSS stations in Svalbard. We computed mass loading for all the GNSS stations in Svalbard and fitted it to the GNSS time series using reciprocal formal uncertainties as weights. Then we computed the $\chi^2$ per 
degree of freedom of the fit and scaled variances of admittance
factor estimates by this amount. Table~\ref{tab:admittance} shows
the estimates of the admittance factor from the differenced GNSS time 
series using Eq.~\ref{eq:ris}. Similar to the VLBI case, the admittance
is very close to 1.0 for the vertical seasonal signal (row ``All'') and
it is far away from 1.0 for the interannual signal and the horizontal signal.

Two factors 
may  cause poor modeling of the interannual signal. First, calving and frontal ablation are not included in 
the \CMB\ model, and therefore, lacks this contribution.
Second, other loadings, for instance non-tidal ocean loading may 
contribute. The seasonal signal has a very specific time dependence 
pattern, and the approach of admittance factor estimation exploits the 
uniqueness of this pattern, while the pattern of the interannual signal
is more general.


\begin{table}
  \caption{Admittance factors for the vertical component of GNSS station in Svalbard caused by 
   the glacier and snow loading. ADM\_SEA and ADM\_IAV show 
     estimates of the seasonal and interannual admittance factors.}
    \label{tab:admittance}
 \par\medskip\par
 \centering
 \begin{tabular}{lrr} \hline
    Station &  ADM\_SEA & ADM\_IAV \\
    \hline
       SVES & 0.94  $\pm$ 0.15 &  0.04 $\pm$ 0.13 \\
       NYAL & 1.33  $\pm$ 0.18 &  0.03 $\pm$ 0.17 \\
       NYA1 & 1.01  $\pm$ 0.18 &  0.19 $\pm$ 0.16 \\
       LYRS & 0.95  $\pm$ 0.18 & -0.11 $\pm$ 0.21 \\
       HORN & 1.35  $\pm$ 0.16 &  0.25 $\pm$ 0.17 \\  
       All  & 1.12  $\pm$ 0.07 &  0.11 $\pm$ 0.07 \\
    \hline
  \end{tabular}
\end{table}

The analysis of the admittance factors give several important results in addition to the very good agreement between the different estimated vertical seasonal components.
Analysis of observations shows that the  \CMB\ model 
provides prediction of the vertical mass loading with 1-$\sigma$ 
errors of 5\%, which corresponds to 0.1~mm.
We have a bias wrt to the model of 0.2  $\pm$ 0.1~mm,
and this bias is not statistically significant at a 95\% level (2 sigma).
 We conclude that analysis of the data from two totally independent techniques, VLBI and GNSS, 
proves there is no statistically significant deviation at a 95\% 
significance level between the seasonal vertical mass loading signal based on the
\CMB\ model and observations of both techniques.

\subsection{Geodynamical interpretation}
\label{ssec:interpretation}
To study the time series ability to capture the loading signal from glaciers and snowpack changes, we used the \CM\ filtered time series. Other known loading signals were removed using Eq.~\ref{eq:ris}.
To have more robust time series, in the following discussion, we have used averaged GNSS time series.
The averaged time series are the weighted mean of the daily values from the Gamit-NMA, the GipsyX-NNR and the GipsyX-FID solutions.
The annual periodic signal and the linear rate for the time series in Eq.~\ref{eq:ris} are included in Table~\ref{tab:sval-up} together with the elastic signal from glaciers and snow. Detailed results for the individual GNSS solutions are included in the Appendix, Table~\ref{tab:sval-up-detailed}.
 
The amplitudes of the estimated loading signal from glaciers and snowpack 
vary with latitude and longitude and depend on the amount of surrounding 
glaciers and land masses (see Fig.~\ref{fig:sval-annual}).
The station HAGN in the middle of the glacier Kongsvegen has the largest 
estimated annual loading signal, while the westernmost stations NYAL/NYA1 
and HORN have the smallest. The GNSS stations SVES and LYRS are located 
in central parts of Svalbard and here the measured vertical annual signal agrees 
with the estimated loading signal at the uncertainty level.  
For the stations closest to the west coast NYAL, NYA1 and HORN the measured 
amplitudes are slightly larger than expected from the variations in 
glaciers and snow ($\sim0.7$~mm, $\sim1.0$~mm and $\sim 0.5$~mm resp.). 
Although the admittance factor for all stations combined show very good 
agreement for the seasonal component, the admittance factor for individual
stations in Ny-\AA lesund and Hornsund are slightly above one, implying that the observed amplitude is somewhat larger than the prediction from the \CMB\ model.

 The larger vertical amplitude at NYAL, NYA1 and HORN might be due to lower precision of the \CMB\ models in areas with more variable coastal climate, changes in groundwater and 
surface hydrology, and seasonal variability in calving/frontal ablation of 
glaciers. Especially, calving is assumed to be seasonally dependent with higher incidents during summer (when ice flows faster). This may explain a higher observed amplitude in areas like Ny-\AA lesund and Hornsund, which have a lot 
of nearby large calving glaciers. However, the deviation of admittance
factors from 1.0 for these stations is still within $2\sigma$ of the 
statistical uncertainty. Longer time series are needed to establish whether 
there is a statistically significant deviation of observations from the model
for these individual stations.

The phase of the vertical loading signal from glaciers and snow varies with only a few days over Svalbard, and corresponds to a maximal value after the end of the melting season, in mid-October. The phase of the GNSS time series agrees with the glaciers and snow signal from the \CMB\ models within a few weeks.

The predicted horizontal seasonal signal is smaller. It is around 0.2~mm in the north component for all locations except HORN. HORN is located in the south of Svalbard with the majority of glaciers located to the north. Consequently the north amplitude is larger, 0.7~mm. The other stations have glaciers both to the north and to the south and the loading signals are cancelled out. The east annual signal varies from 0.7~mm (NYAL, NYA1) to 0.0~mm (SVES), depending on their locations relative to the glaciers.

We see that the estimated annual signal for the GNSS stations NYA1, HAGN, LYRS and HORN agree at 2-sigma level in the east component. HORN is capturing the larger north annual signal for this location. However, the small north signals for the other stations are too small to be detected. LYRS and SVES have a large north annual signal, 1.1~mm resp. 2.3~mm larger than expected loading signal. We have not established the origin of this discrepancies. Possible reasons are uneven thermal expansion of the antenna monument (steel mast) and artifacts of the atmosphere model. We will investigate these signals in the future.

Our CMB model is limited by Svalbard archipelago. Glacial loading at
other islands, such as Iceland and Greenland, can bring a noticeable 
contribution. Using Green's function from the ocean tide loading 
provide of \citet{r:Scherneck1991}\footnote{Fig.~2 in 
http://holt.oso.chalmers.se/loading/loadingprimer.html}, assuming 100~Gt 
seasonal ice cycle on Greenland \citep[see figure~6 in][]{r:Bevis2012}, 
and the average distance from Greenland to Svalbard of 800~km, we get 
coarse estimate of the the amplitude of mass loading signal Svalbard due to
glacier in Greenland: 0.38~mm. In order to get a more refined estimate of 
the magnitude of such a contribution, we used \citet{r:loo19} mascon 
solution for Greenland for 2008 from processing GRACE mission. The mascon 
for Greenland at a regular grid $0.5^\circ \times 0.5^\circ$ is 
provided\footnote{Available at https://earth.gsfc.nasa.gov/geo/data/grace-mascons} 
with a monthly resolution in the height of the column of water 
equivalent that covers the entire land and is zero otherwise. The mascon 
excludes the contribution of the atmosphere and ocean but retains the 
contribution of land water, snow, and ice storage.

\begin{figure}
   \includegraphics[width=0.490\textwidth]{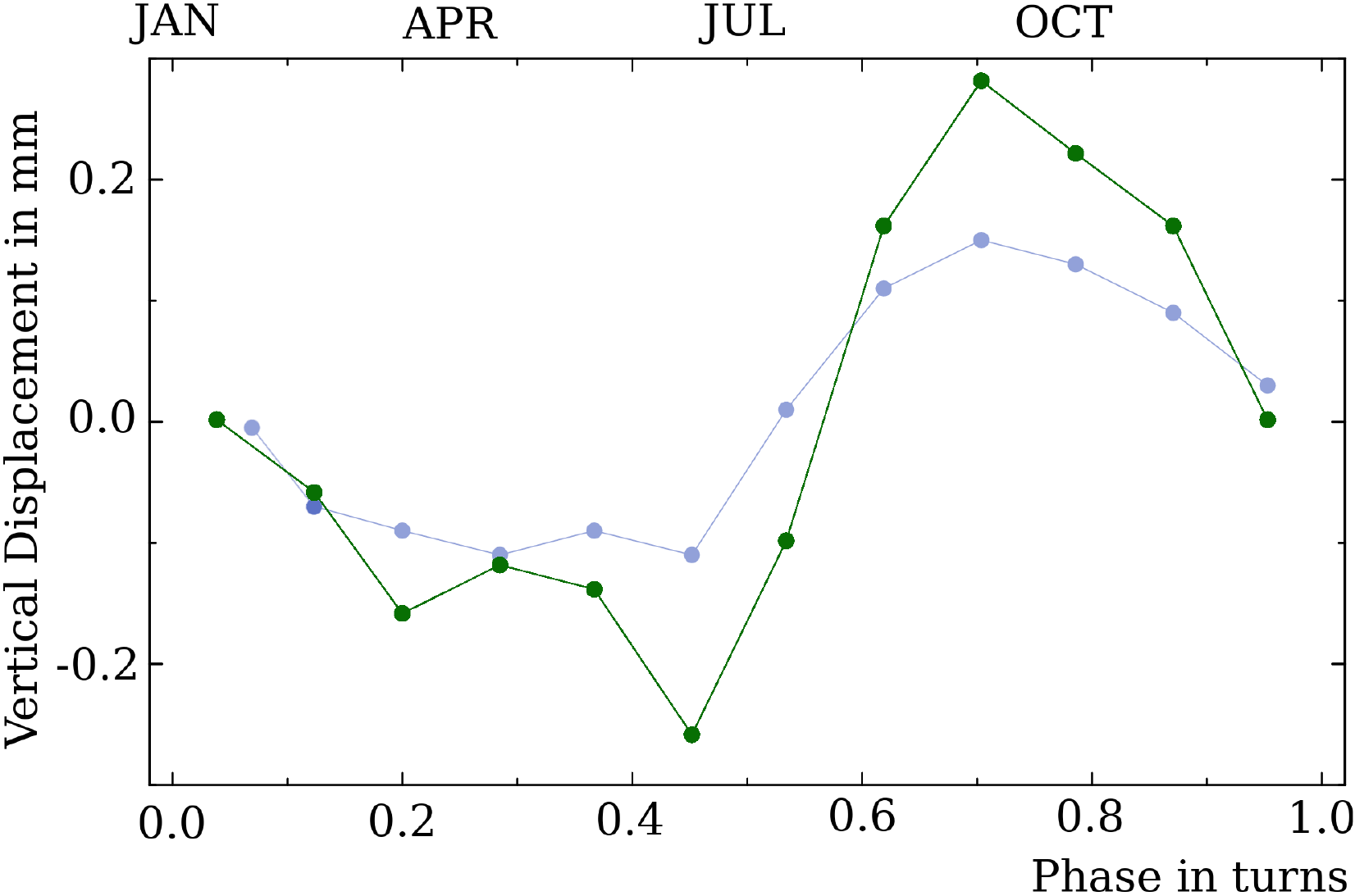}
   \caption{The seasonal contribution of glaciers in Greenland to 
            the vertical displacement of NYALES20 after removal of the 
            slowly varying constituent from processing GRACE data. 
            The green line shows the total contribution, the shadowed
            blue line shows the residual contribution after subtraction
            of land water storage pressure from MERRA2 model. The horizontal
            axis is a phase of the seasonal signal in phase turns.}
\label{fig:greenland}
\end{figure}

\begin{figure}
\centering
  \includegraphics[width=.7\textwidth]{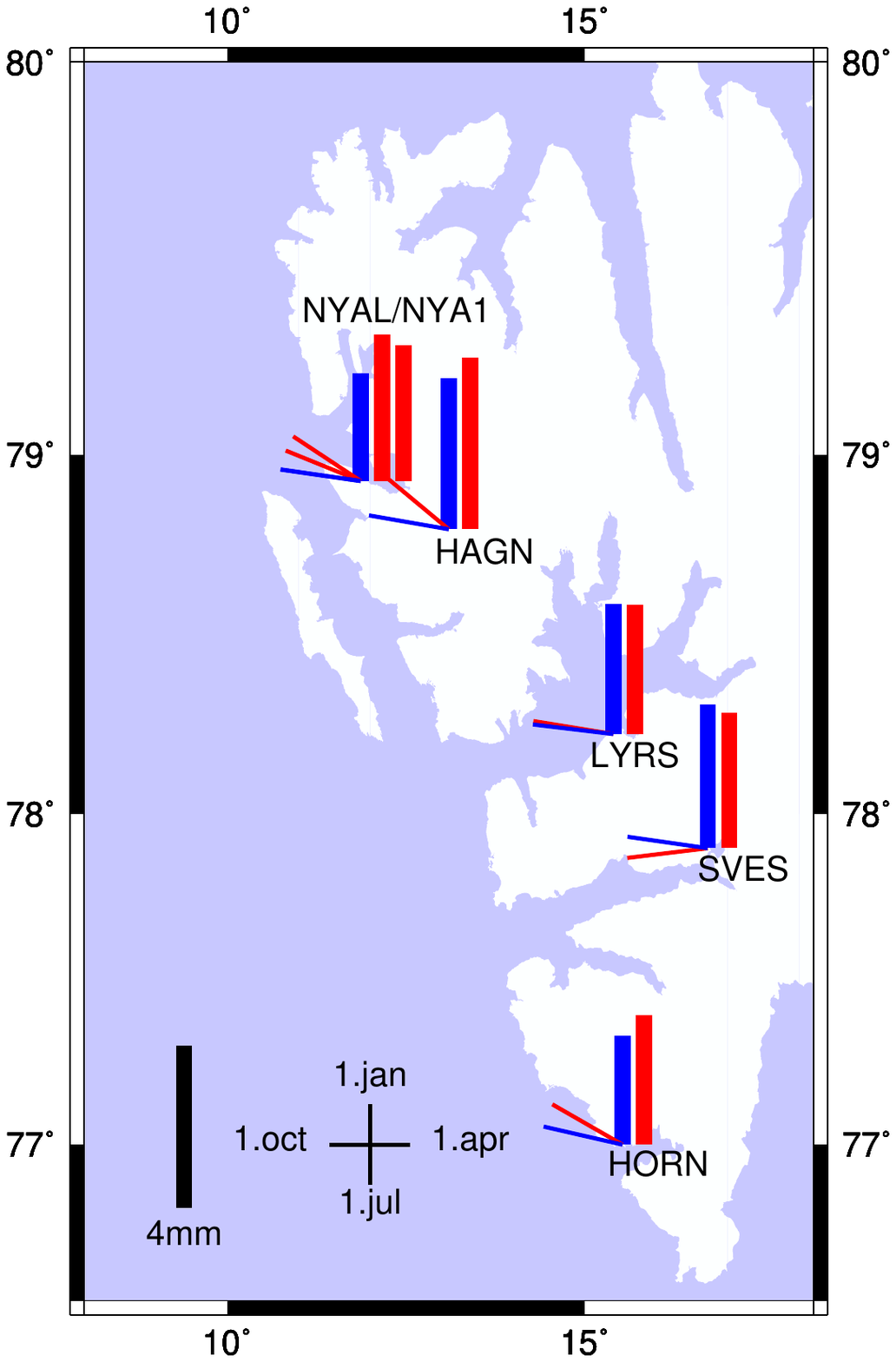}
\caption{Annual signal for GNSS stations in Svalbard. The bars are the amplitude and the vectors are the phase. Blue is from the loading prediction from glaciers and snow red is from the GNSS stations.}
\label{fig:sval-annual}       
\end{figure}

We have converted the height of the column of water equivalent to 
surface pressure and computed the mass loading from the mascon using
the same approach as we used for computing atmospheric and land water 
storage loading using spherical harmonic transform of degree/order 2699. 
Since MERRA2 land water storage model covers Greenland, we computed 
mass loading two times: the first time using the total surface pressure 
from the mascon and the second time after subtracting the pressure anomaly 
from MERRA2. In the latter case the resulting mass loading signal provides 
a correction to the mass loading from MERRA2 model for the contribution 
of glacier derived from GRACE data analysis since MERRA2 has large errors 
of modeling glacier dynamics. The results are shown in Figure~\ref{fig:greenland} after removal linear trend. The residual signal has amplitude 0.15~mm
and is in phase with the mass loading signal from Svalbard glacier
while the total signal has amplitude 0.28~mm. This estimate agrees
remarkably well with our coarse estimate. Accounting for the contribution
of glacier mass loading using GRACE data reduces the admittance factor 
from 1.10 to 1.04 for the seasonal vertical component of the VLBI solution.

We exercise a caution in results of processing GRACE data,
A thorough analysis of systematic errors of mass loading signal
from GRACE mascon solution requires significant efforts and is beyond
the scope of the present manuscript. However, our estimates
shows that the contribution of glaciers in Greenland is not dominated
and its accounting improves the agreement of the CMB model with VLBI
and GNSS observations.

\begin{table}
\caption{Vertical rate and annual signal for GNSS stations in Svalbard. GS are the elastic loading signal from ice and snow. GNSS-CM is the time series using Eq.~\ref{eq:ris}. Max uplift is the date of the maximum value for the annual signal.}
\label{tab:sval-up}
\centering
\scriptsize
\begin{tabular}{llrrrrr}
  \hline
  Station & & Trend \hfill   & Amp. \hfill & Pha.  & Max up \\ 
          & & mm/yr        & mm \hfill & deg & date         \\
  \hline
Up:\\
 NYA1 &     GNSS-CML &  9.74 +/-  0.27 &   3.37 +/-   0.29 &  -56.5 +/-    5.0 &   3 Nov.   \\
 &         GS &  0.93 +/-  0.03 &   2.66 +/-   0.03 &  -81.8 +/-    0.6 &   9 Oct.   \\
NYAL &     GNSS-CML &  9.57 +/-  0.27 &   3.63 +/-   0.29 &  -67.9 +/-    4.6 &  23 Oct.   \\
 &         GS &  0.93 +/-  0.03 &   2.66 +/-   0.03 &  -81.8 +/-    0.6 &   9 Oct.   \\
HAGN &     GNSS-CML & 11.95 +/-  0.56 &   4.28 +/-   0.65 &  -50.2 +/-    8.6 &  10 Nov.   \\
 &         GS &  1.81 +/-  0.04 &   3.73 +/-   0.04 &  -80.1 +/-    0.7 &  10 Oct.   \\
LYRS &     GNSS-CML &  8.16 +/-  0.42 &   3.21 +/-   0.45 &  -80.8 +/-    8.0 &  10 Oct.   \\
 &         GS &  0.83 +/-  0.03 &   3.21 +/-   0.03 &  -83.2 +/-    0.6 &   7 Oct.   \\
SVES &     GNSS-CML &  6.21 +/-  0.45 &   3.37 +/-   0.47 &  -96.8 +/-    7.9 &  23 Sep.   \\
 &         GS &  0.86 +/-  0.04 &   3.53 +/-   0.04 &  -81.9 +/-    0.6 &   8 Oct.   \\
HORN &     GNSS-CML &  9.45 +/-  0.27 &   3.21 +/-   0.30 &  -60.0 +/-    5.3 &  31 Oct.   \\
 &         GS &  1.93 +/-  0.03 &   2.69 +/-   0.03 &  -77.0 +/-    0.6 &  13 Oct.   \\
North:\\
NYA1 &     GNSS-CML & 14.98 +/-  0.09 &   0.24 +/-   0.09 &   39.6 +/-   21.1 &   9 Feb.   \\
 &         GS &  0.56 +/-  0.01 &   0.17 +/-   0.01 &  -72.0 +/-    1.0 &  18 Oct.   \\
NYAL &     GNSS-CML & 14.84 +/-  0.12 &   0.47 +/-   0.12 &  -89.5 +/-   14.9 &   1 Oct.   \\
 &         GS &  0.56 +/-  0.01 &   0.17 +/-   0.01 &  -72.0 +/-    1.0 &  18 Oct.   \\
HAGN &     GNSS-CML & 14.73 +/-  0.36 &   0.83 +/-   0.34 &  147.7 +/-   22.2 &  29 May.   \\
 &         GS &  0.53 +/-  0.01 &   0.05 +/-   0.01 &  -49.1 +/-    3.4 &  11 Nov.   \\
LYRS &     GNSS-CML & 14.47 +/-  0.16 &   1.19 +/-   0.17 &   63.6 +/-    8.2 &   5 Mar.   \\
 &         GS &  0.24 +/-  0.01 &   0.06 +/-   0.01 &   84.4 +/-    3.6 &  26 Mar.   \\
SVES &     GNSS-CML & 14.49 +/-  0.33 &   2.55 +/-   0.33 &   35.4 +/-    7.3 &   4 Feb.   \\
 &         GS &  0.19 +/-  0.01 &   0.24 +/-   0.01 &   89.0 +/-    1.2 &  31 Mar.   \\
HORN &     GNSS-CML & 13.20 +/-  0.14 &   1.02 +/-   0.14 &   91.7 +/-    8.0 &   2 Apr.   \\
 &         GS & -0.43 +/-  0.01 &   0.73 +/-   0.01 &  101.6 +/-    0.7 &  13 Apr.   \\
East:\\
NYA1 &     GNSS-CML & 10.24 +/-  0.08 &   0.42 +/-   0.09 &   18.6 +/-   11.8 &  18 Jan.   \\
 &         GS & -0.07 +/-  0.01 &   0.64 +/-   0.01 &   98.3 +/-    0.8 &   9 Apr.   \\
NYAL &     GNSS-CML & 10.01 +/-  0.08 &   0.16 +/-   0.07 &  -59.1 +/-   25.0 &   1 Nov.   \\
 &         GS & -0.07 +/-  0.01 &   0.64 +/-   0.01 &   98.3 +/-    0.8 &   9 Apr.   \\
HAGN &     GNSS-CML & 12.65 +/-  0.32 &   0.66 +/-   0.29 &  151.4 +/-   23.9 &   2 Jun.   \\
 &         GS &  0.32 +/-  0.01 &   0.40 +/-   0.01 &   94.5 +/-    0.9 &   5 Apr.   \\
LYRS &     GNSS-CML & 12.48 +/-  0.17 &   0.35 +/-   0.15 &   77.7 +/-   23.5 &  19 Mar.   \\
 &         GS &  0.08 +/-  0.01 &   0.27 +/-   0.01 &   95.9 +/-    1.1 &   7 Apr.   \\
SVES &     GNSS-CML & 15.71 +/-  0.27 &   0.76 +/-   0.26 &   31.9 +/-   18.8 &   1 Feb.   \\
 &         GS & -0.17 +/-  0.01 &   0.07 +/-   0.01 &  117.7 +/-    2.5 &  29 Apr.   \\
HORN &     GNSS-CML & 11.56 +/-  0.11 &   0.45 +/-   0.11 &   51.2 +/-   13.8 &  20 Feb.   \\
 &         GS & -0.31 +/-  0.01 &   0.32 +/-   0.01 &  106.4 +/-    0.9 &  17 Apr.   \\
\hline 
\end{tabular}
\end{table}

\subsection{CMB model and time series}
\clearpage
\label{ssec:rishyd}
In the previous section we examined how the different GNSS time series were able to capture the elastic loading signal from local ice and snow changes. In this paragraph we will discuss the effect of removing the loading signal from the time series, both on the unfiltered time series and the \CM\ filtered time series. We will in particular look at the effect of replacing the global hydrological model with a regional \CMB\ model. In the discussion we used an averaged time series from the GNSS solutions; Gamit-NMA, GipsyX-FID and GipsyX-NNR. Due to limited observations during winter the HAGN time series are not directly comparable with the other time series and therefore not included in this discussion.

 We have used time series where no loading models where removed and two time series with slightly different loading time series removed ($L_1$, $L_2$).
The first loading model, $L_1$, is the sum of the loadings from \ATM, \NTO\ and the total \LWS\ signal from the merra2 model. The second loading model, $L_2$, equlas $L_1$ except that the regional \LWS\ signal in merra2 is replaced with the glacier and snow signal in the \CMB\ model using Eq.~\ref{eq:merra_modi}.

$H_{GNSS}-L_i$ contains the unmodeled signal in the GNSS time series after removing the modeled loadings, $L_i$. The signal can be presented as a sum of linear trend (from e.g., GIA and tectonics) and noise including unmodeled loading signals. I.e. $H_{GNSS}(t)-L_i(t) = LIN(t)+\varepsilon(t)$. Also the \CM\ filtered time series ($H_{CM}$, Eq.~\ref{eq:cm}) and the \CM\ filtered time series where the loading signal is removed ($H_{CM,L_i}$, Eq.~\ref{eq:cm-time-series}) can be presented as a sum $LIN(t)+\varepsilon(t)$. To examine the quality of the time series using different filtering and loading models we estimate the \RMS\ and annual signal in the noise time series $\varepsilon(t)$. The annual signal in $\varepsilon(t)$ is the remaining annual signal after removing the loading model $L_i$. A large annual signal in $\varepsilon(t)$ indicate that we have remaining unmodelled periodic signal in the time series after removing load $L_i$.  The \RMS\ is a measure of the remaining noise in the time series. The results for the averaged time series are included in Table~\ref{tab:sval-quality}. Results for the individual solutions are included in Table~\ref{tab:sval-quality-detailed} in the Appendix.

\begin{table*}
\caption{Yearly amplitude and RMS in the time series. In each column the three parameters are amplitude of yearly signal in mm, \RMS\ of the time series in mm and changes in \RMS\ relative to the unfiltered time series in percent. The numbers are in mm. $H_{GNSS}$, $L_i$ and $H_{CM,L_i}$ are explained in the text.} 
\label{tab:sval-quality}
\scriptsize{
\centering
\begin{tabular}{lrrrrrr}\hline
  Station & $H_{GNSS}$ & $H_{GNSS}-L_1$ &  $H_{GNSS}-L_2$& $H_{CM}$ & $H_{CM,L_1}$ &  $H_{CM,L_2}$ \\ \hline
 BJOS&  1.3 / 4.9(   0\%)&  2.9 / 4.1( -16\%)&  3.0 / 4.1( -16\%)&  \\
 NYA1&  3.8 / 4.6(   0\%)&  3.7 / 4.0( -13\%)&  3.7 / 3.9( -16\%)&  3.4 / 4.7(   2\%)&  2.5 / 4.5(  -3\%)&  1.5 / 4.4(  -4\%) \\
 NYAL&  3.6 / 4.6(   0\%)&  3.1 / 4.1( -12\%)&  3.0 / 3.9( -15\%)&  3.4 / 4.7(   3\%)&  2.5 / 4.5(  -3\%)&  1.3 / 4.4(  -4\%) \\
 LYRS&  3.0 / 6.8(   0\%)&  2.4 / 6.3(  -8\%)&  2.9 / 6.2( -10\%)&  2.9 / 6.7(  -1\%)&  1.7 / 6.6(  -3\%)&  0.6 / 6.6(  -4\%) \\
 SVES&  2.5 / 6.5(   0\%)&  1.7 / 5.8(  -9\%)&  2.8 / 5.7( -11\%)&  2.9 / 6.3(  -2\%)&  1.5 / 6.1(  -5\%)&  1.1 / 6.0(  -7\%) \\
 HORN&  3.3 / 4.7(   0\%)&  3.3 / 4.1( -14\%)&  3.4 / 3.8( -20\%)&  3.0 / 4.7(  -0\%)&  2.4 / 4.5(  -5\%)&  1.0 / 4.3(  -9\%) \\
\hline
\end{tabular}}
\end{table*}

We see that removal of the loading signal reduce the \RMS\ values on average by 11\%, while replacing the regional hydrological signal with a \CMB\ model reduce the \RMS\ with 13\%. The improvements for the \CM\ filtered time series are less, 4\% and 6\%, respectively. The removal of the \CM\ eliminates part of the elastic loading signal, and this may explains the lower reduction for these series. Both for the unfiltered and the \CM\ filtered time series the \RMS\ are reduced with 2--3 \% when we replace the regional signal in the merra2 with the glacier and snow signal from the \CMB\ model.    
The \RMSs\ are very little affected by the \CM\ filtering (4th vers. 1st column in Table~\ref{tab:sval-quality}).

Removing the \NTL\ (\ATM, \NTO\ and \LWS\ including the glaciers and snow) from the observed time series have an effect on the daily noise scatter (\RMS), but  very little effect on the annual signal. This implies that removal of the \NTL\ reduces the daily scatter in the GNSS time series. It also implies that the periodic signal is dominated by other factors. As we saw in Section~\ref{sec:results} this annual signal depends on the analysis strategy. We conclude that we have an analysis strategy dependent effect in the periodic signal.

The amplitude of the time series is reduced after the \CM\ filtering. The amplitude of the annual signal in the \CM\ filtered time series using load models $L_1$ is reduced to $2.1$~mm. The largest effects are when we use load model $L_2$ and the \CM\ filtered time series. For this solution the averaged annual loading signal is $1.1$~mm, one third of most other combinations of filtering and loading models.

For the horizontal components including \CM\ filtering and removing the NTL 
affects the RMS only marginaly. The amplitude of the horizontal seasonal signal reduces from 
on average 0.6~mm to 0.4~mm for NYAL, NYA1 and HORN when both CM filtering 
and NTL loadings are removed. The stations LYRS and SVES have much larger amplitudes and 
no improvements after filtering. 

Note, the glacier model used in this study is not able to capture glacier dynamics like continuous flow of ice towards the glacier front, or more dramatic phenomena such as glacier surging \citep[see e.g.,][]{morris++2020,dunse++2015}. These dynamic effects provide a significant contribution to the total glacier mass balance and uplift, especially, on time scales from years and longer \citep[see][for more on the effect of glacier dynamics on the uplift]{kierulf++2009a}. The linear elastic uplift signal from the \CMB\ models is not sufficient to fully describe the elastic uplift from ice and snow changes over longer time scales.

\section{Conclusions}
\label{sec:conclusions}
In the introduction two questions were asked: (1) How well do GNSS and VLBI capture the seasonal loading signal from glaciers and snow on Svalbard? (2) Will refining the \LWS\ models with a \CMB\ model improve the loading predictions? To answer these questions a network of seven permanent GNSS stations were analyzed with different analysis strategies and softwares. The different time series were studied and compared with loading predictions from \ATM, \NTO, \LWS\ including glaciers and snow.

We found large discrepancies between the different analysis strategies, 
both in phase and amplitude, while the estimates of long-term trend 
were more consistent. This implies that a direct geophysical interpretation of 
raw GNSS time series is problematic. To overcome this problem,
we performed \CM\ filtering utilizing the data from the nearby 
station at Bear Island. The elastic loading signal was removed from the time series before the \CM\ filtering. The \CM\ filtered time series gave a much better agreement.
This confirmed our initial conjecture that the origin of the discrepancies 
in the raw time series are due to differences in the analysis strategy in the GNSS data processing. The agreement of \CM\ filtered time series 
strengthened our confidence that we investigated a real geophysical 
signal, and not artifacts of data analysis.

We have decomposed the \LWS\ signal into the seasonal
and interannual signals, and estimated admittance factors from VLBI
data and GNSS \CM\ filtered time series. The admittance factors estimates 
from vertical seasonal constituents for both techniques do not deviate 
from 1.0 at a $2\sigma$ level. Therefore, we conclude that 
the entire mass loading signal is present in data from totally 
independent technique at a statistical significance level of 95\%. 
The $1\sigma$ uncertainty of admittance factors corresponds to 0.1~mm.
This implies that by using the CMB model we can predict seasonal vertical mass 
loading displacements on Svalbard with the same level of accuracy, and that predicted errors are less than the observation errors. The interannual 
loading signal was not recovered from observations. Further work is
required to explain these discrepancies. However, since calving and frontal ablation are not included in the \CMB\ model, they may contribute to these discrepancies.

We saw a significant reduction of residuals 
after subtraction of the \LWS\ mass loading displacement.
The annual amplitude was reduced from $2.1$~mm to $1.1$~mm 
in the \CM\ filtered time series. Subtraction of the \LWS\ loading had a
negligible impact on the unfiltered time series. This provides a strong evidence that
\CM\ filtering is necessary to 
reveal local periodic signals when millimeter accuracy is required.

\section*{Acknowledgments}
We thank the editor Prof. Bert Vermeersen,  Shfaqat Abbas Khan and one anonymous reviewer for their constructive comments that helped to improve the manuscript.
Thanks to Zuheir Altamimi for providing the \ITRF\ time series for the Ny-\AA lesund stations. Thanks to Machiel Bos for providing the Hector software and including new features utilized in this study. We thank Bryant Loomis for providing GRACE mascon results and for a fruitful discussion. Ward van Pelt acknowledges funds from the Swedish National Space Agency (project 189/18).

\section*{Data availability}
The GNSS data from Hornsund (HORN) is provided by the Institute of Geophysics of the Polish Academy of Sciences (IG PAS). Time series from SOPAC, JPL and UNR were used in the study. GNSS time series and SCG data is available from the authors. The NTL time series are avialable from \web{http://massloading.net}. Distributed time series of climatic mass balance were extracted from the dataset described in \cite{vanpelt++2019}.

\bibliographystyle{gji}
\newcommand{\newblock}{}
\bibliography{svalbard}

\clearpage

\appendix
\section{Tables}

\begin{table*}
\caption{Trend and annual signal in Ny-\AA lesund and Bear Island. The parameters 
are estimated trend and annual signal estimated using Eq.~\ref{eq:trend}) for the North and East components.}
\label{tab:nyal-horizontal}
\footnotesize{
\centering
\begin{tabular}{llrrrrrrr}\hline
  Station & & \multicolumn {3}{c}{North}& \multicolumn {3}{c}{East}            \\ 
          & &  Trend  & Amp. & Pha. & Trend & Amp. & Pha.  \\ 
          & & mm/yr & mm & deg & mm/yr &  mm & deg   \\ 
  \hline
NYA1 & Gamit-SOPAC & 14.83 +/-  0.12 &   0.26 +/-   0.11  &  -11.5 +/-   22.7  & 10.27 +/-  0.12 &   1.04 +/-   0.12  &   -6.0 +/-    6.7  \\
NYA1 &  Gamit-NMA & 14.97 +/-  0.11 &   0.44 +/-   0.11  &  -61.8 +/-   14.0  & 10.46 +/-  0.10 &   0.61 +/-   0.10  &  -33.2 +/-    9.6  \\
NYA1 & GipsyX-FID & 15.00 +/-  0.13 &   0.67 +/-   0.13  & -121.4 +/-   11.4  & 10.19 +/-  0.12 &   1.19 +/-   0.12  &    6.4 +/-    5.8  \\
NYA1 & GipsyX-NNR & 15.01 +/-  0.13 &   0.63 +/-   0.13  & -116.2 +/-   11.6  & 10.20 +/-  0.11 &   1.08 +/-   0.12  &   15.1 +/-    6.1  \\
NYA1 & GipsyX-UNR & 14.93 +/-  0.13 &   0.64 +/-   0.13  & -113.7 +/-   11.6  & 10.30 +/-  0.11 &   1.06 +/-   0.12  &   16.7 +/-    6.4  \\
NYA1 & GipsyX-JPL & 14.67 +/-  0.14 &   0.78 +/-   0.14  & -108.4 +/-   10.2  & 10.31 +/-  0.13 &   0.85 +/-   0.13  &   27.7 +/-    8.9  \\
NYA1 &   ITRF2014 & 14.46 +/-  0.17 &   0.73 +/-   0.14  &  -83.2 +/-   10.7  & 10.40 +/-  0.18 &   1.02 +/-   0.15  &    2.8 +/-    8.2  \\
NYA1 &  Gamit-NMA (CM)& 14.97 +/-  0.09 &   0.44 +/-   0.09  &   20.1 +/-   11.9  & 10.50 +/-  0.10 &   0.41 +/-   0.10  &  -37.2 +/-   14.1  \\
NYA1 & GipsyX-FID (CM) & 14.98 +/-  0.10 &   0.29 +/-   0.10  &   18.1 +/-   19.3  & 10.23 +/-  0.09 &   0.41 +/-   0.09  &   12.7 +/-   13.0  \\
NYA1 & GipsyX-NNR (CM) & 15.00 +/-  0.09 &   0.34 +/-   0.10  &   31.1 +/-   16.1  & 10.24 +/-  0.08 &   0.38 +/-   0.09  &   18.4 +/-   12.7  \\
\hline 
NYAL & Gamit-SOPAC & 14.79 +/-  0.12 &   0.84 +/-   0.13  &  -79.5 +/-    8.8  &  9.97 +/-  0.11 &   0.47 +/-   0.12  &  -21.8 +/-   13.8  \\
NYAL &  Gamit-NMA & 14.91 +/-  0.12 &   1.12 +/-   0.13  &  -85.6 +/-    6.6  & 10.14 +/-  0.11 &   0.51 +/-   0.11  &  -95.4 +/-   12.3  \\
NYAL & GipsyX-FID & 14.86 +/-  0.15 &   1.20 +/-   0.17  & -114.5 +/-    7.9  &  9.97 +/-  0.13 &   0.90 +/-   0.15  &  -14.2 +/-    9.3  \\
NYAL & GipsyX-NNR & 14.87 +/-  0.14 &   1.19 +/-   0.15  & -111.8 +/-    7.1  &  9.99 +/-  0.13 &   0.74 +/-   0.13  &    0.8 +/-   10.2  \\
NYAL & GipsyX-UNR & 14.86 +/-  0.14 &   1.17 +/-   0.15  & -110.6 +/-    7.3  & 10.00 +/-  0.13 &   0.75 +/-   0.13  &    0.9 +/-   10.2  \\
NYAL & GipsyX-JPL & 14.62 +/-  0.16 &   1.33 +/-   0.16  & -110.6 +/-    6.9  & 10.00 +/-  0.13 &   0.48 +/-   0.13  &   12.5 +/-   15.5  \\
NYAL &   ITRF2014 & 14.55 +/-  0.21 &   1.20 +/-   0.17  &  -89.9 +/-    7.9  & 10.30 +/-  0.19 &   0.62 +/-   0.15  &    0.3 +/-   13.4  \\
NYAL &  Gamit-NMA (CM) & 14.91 +/-  0.11 &   0.68 +/-   0.11  &  -64.3 +/-    9.5  & 10.18 +/-  0.10 &   0.48 +/-   0.10  & -121.8 +/-   12.2  \\
NYAL & GipsyX-FID (CM) & 14.84 +/-  0.13 &   0.50 +/-   0.14  &  -65.2 +/-   15.4  & 10.01 +/-  0.09 &   0.29 +/-   0.10  &  -72.2 +/-   18.7  \\
NYAL & GipsyX-NNR (CM) & 14.85 +/-  0.12 &   0.40 +/-   0.12  &  -71.0 +/-   16.6  & 10.02 +/-  0.09 &   0.19 +/-   0.08  &  -72.3 +/-   23.3  \\
\hline 
Ny-\AA lesund &        NTL &  0.56 +/-  0.04 &   0.35 +/-   0.04  & -127.3 +/-    7.0  & -0.04 +/-  0.06 &   0.94 +/-   0.06  &   59.2 +/-    3.8  \\
\hline 
BJOS &  Gamit-NMA & 13.37 +/-  0.11 &   0.55 +/-   0.12  & -110.7 +/-   11.8  & 13.74 +/-  0.13 &   0.29 +/-   0.12  &  -32.7 +/-   22.3  \\
BJOS & GipsyX-FID & 13.30 +/-  0.16 &   0.94 +/-   0.17  & -134.2 +/-   10.3  & 13.57 +/-  0.15 &   0.84 +/-   0.16  &    3.6 +/-   10.6  \\
BJOS & GipsyX-NNR & 13.31 +/-  0.15 &   0.94 +/-   0.16  & -128.1 +/-    9.7  & 13.58 +/-  0.15 &   0.74 +/-   0.15  &   17.2 +/-   11.5  \\
\hline
Bear Island &        NTL & -0.06 +/-  0.05 &   0.45 +/-   0.05  & -163.0 +/-    6.8  &  0.07 +/-  0.06 &   0.75 +/-   0.06  &   20.3 +/-    4.7  \\
\hline
\end{tabular}
}
\end{table*}

\begin{table*}
\caption{NTL vertical variations at GNSS stations in Svalbard.}
\label{tab:sval-loading-detailed}
\centering
\small{
\begin{tabular}{llrrrr}\hline
  Station & & Trend   & Amp. & Pha. & Max up  \\ 
          & & mm/yr & mm & deg  & date \\ 
\hline 
NYA1 &        ATM &  0.14 +/-  0.14 &   1.02 +/-   0.14 &   -4.1 +/-     8.0 &  26 Dec.  \\ 
     &        NTO & -0.06 +/-  0.23 &   0.44 +/-   0.20 &  167.2 +/-    24.5 &  18 Jun.  \\ 
     &        HYD & -0.11 +/-  0.01 &   1.40 +/-   0.01 & -111.8 +/-     0.3 &   8 Sep.  \\ 
     &   Snowpack & -0.16 +/-  0.01 &   0.83 +/-   0.01 &  -89.8 +/-     0.6 &  30 Sep.  \\ 
     &        Ice &  1.09 +/-  0.02 &   1.84 +/-   0.02 &  -78.3 +/-     0.7 &  12 Oct.  \\ 
     & Total loads &  0.92 +/-  0.30 &   4.00 +/-   0.32 &  -82.5 +/-     4.5 &   8 Oct.  \\ 
\hline 
NYAL &        ATM &  0.14 +/-  0.14 &   1.02 +/-   0.14 &   -4.1 +/-     8.0 &  26 Dec.  \\ 
     &        NTO & -0.06 +/-  0.23 &   0.44 +/-   0.20 &  167.2 +/-    24.5 &  18 Jun.  \\ 
     &        HYD & -0.11 +/-  0.01 &   1.40 +/-   0.01 & -111.8 +/-     0.3 &   8 Sep.  \\ 
     &   Snowpack & -0.16 +/-  0.01 &   0.83 +/-   0.01 &  -89.8 +/-     0.6 &  30 Sep.  \\ 
     &        Ice &  1.09 +/-  0.02 &   1.84 +/-   0.02 &  -78.3 +/-     0.7 &  12 Oct.  \\ 
     & Total loads &  0.92 +/-  0.30 &   4.00 +/-   0.32 &  -82.5 +/-     4.5 &   8 Oct.  \\ 
\hline 
HAGN &        ATM &  0.15 +/-  0.16 &   1.12 +/-   0.17 &   -3.0 +/-     8.4 &  28 Dec.  \\ 
     &        NTO & -0.07 +/-  0.22 &   0.44 +/-   0.20 &  167.7 +/-    24.0 &  19 Jun.  \\ 
     &        HYD & -0.10 +/-  0.01 &   1.41 +/-   0.01 & -111.8 +/-     0.3 &   8 Sep.  \\ 
     &   Snowpack & -0.24 +/-  0.01 &   0.81 +/-   0.01 &  -88.3 +/-     0.6 &   2 Oct.  \\ 
     &        Ice &  2.05 +/-  0.04 &   2.92 +/-   0.04 &  -77.7 +/-     0.7 &  13 Oct.  \\ 
     & Total loads &  1.80 +/-  0.32 &   5.06 +/-   0.33 &  -80.1 +/-     3.7 &  10 Oct.  \\ 
\hline 
LYRS &        ATM &  0.15 +/-  0.15 &   1.02 +/-   0.16 &   -3.9 +/-     8.9 &  27 Dec.  \\ 
     &        NTO & -0.09 +/-  0.22 &   0.46 +/-   0.20 &  169.1 +/-    23.6 &  20 Jun.  \\ 
     &        HYD & -0.10 +/-  0.01 &   1.45 +/-   0.01 & -111.9 +/-     0.4 &   8 Sep.  \\ 
     &   Snowpack & -0.06 +/-  0.01 &   1.35 +/-   0.01 &  -89.7 +/-     0.6 &   1 Oct.  \\ 
     &        Ice &  0.89 +/-  0.02 &   1.87 +/-   0.02 &  -78.5 +/-     0.7 &  12 Oct.  \\ 
     & Total loads &  0.80 +/-  0.31 &   4.57 +/-   0.33 &  -84.4 +/-     4.1 &   6 Oct.  \\ 
\hline 
SVES &        ATM &  0.15 +/-  0.17 &   1.04 +/-   0.17 &   -3.5 +/-     9.2 &  27 Dec.  \\ 
     &        NTO & -0.10 +/-  0.22 &   0.47 +/-   0.20 &  169.3 +/-    23.3 &  20 Jun.  \\ 
     &        HYD & -0.10 +/-  0.01 &   1.48 +/-   0.01 & -112.0 +/-     0.3 &   8 Sep.  \\ 
     &   Snowpack & -0.10 +/-  0.01 &   1.13 +/-   0.01 &  -89.2 +/-     0.6 &   1 Oct.  \\ 
     &        Ice &  0.95 +/-  0.03 &   2.41 +/-   0.03 &  -78.5 +/-     0.7 &  12 Oct.  \\ 
     & Total loads &  0.81 +/-  0.33 &   4.92 +/-   0.34 &  -83.3 +/-     3.9 &   7 Oct.  \\ 
\hline 
HORN &        ATM &  0.13 +/-  0.13 &   0.86 +/-   0.13 &   -3.5 +/-     8.7 &  27 Dec.  \\ 
     &        NTO & -0.10 +/-  0.23 &   0.48 +/-   0.21 &  168.9 +/-    23.7 &  20 Jun.  \\ 
     &        HYD & -0.10 +/-  0.01 &   1.50 +/-   0.01 & -112.0 +/-     0.3 &   8 Sep.  \\ 
     &   Snowpack & -0.02 +/-  0.01 &   0.50 +/-   0.01 &  -89.6 +/-     0.7 &   1 Oct.  \\ 
     &        Ice &  1.95 +/-  0.03 &   2.20 +/-   0.03 &  -74.1 +/-     0.7 &  16 Oct.  \\ 
     & Total loads &  1.87 +/-  0.31 &   4.05 +/-   0.32 &  -82.8 +/-     4.5 &   8 Oct.  \\ 
\hline
\end{tabular}}
\end{table*}

\clearpage
\begin{table*}
\caption{Vertical rate and annual signal for GNSS stations in Svalbard. The results are based on the time series using Eq.~\ref{eq:ris}. Max uplift is the date of the maximum value for the annual signal.}
\label{tab:sval-up-detailed}
\centering
\small{
\begin{tabular}{llrrrrr}\hline
  Station & & Trend   & Amp. & Pha. & Max uplift    \\ 
  & & mm/yr & mm & deg & date \\
  \hline
NYA1 & Gamit-NMA-CML &  9.51 +/-  0.37 &   4.14 +/-   0.39 &  -59.6 +/-    5.4  &  31 Oct.   \\
     & GipsyX-FID-CML &  9.75 +/-  0.25 &   3.03 +/-   0.28 &  -71.2 +/-    5.2  &  19 Oct.   \\
     & GipsyX-NNR-CML &  9.82 +/-  0.28 &   3.20 +/-   0.30 &  -72.8 +/-    5.4 &  18 Oct.   \\
     &     GNSS-CML &  9.74 +/-  0.27 &   3.37 +/-   0.29 &  -56.5 +/-    5.0 &    3 Nov.   \\
\hline 
NYAL & Gamit-NMA-CML &  9.46 +/-  0.32 &   3.74 +/-   0.34 &  -65.5 +/-    5.1 &   25 Oct.   \\
     & GipsyX-FID-CML &  9.62 +/-  0.24 &   3.60 +/-   0.29 &  -77.4 +/-    4.7 &   13 Oct.   \\
     & GipsyX-NNR-CML &  9.69 +/-  0.27 &   3.80 +/-   0.29 &  -78.6 +/-    4.4 &    12 Oct.   \\
     &     GNSS-CML &  9.57 +/-  0.27 &   3.63 +/-   0.29 &  -67.9 +/-    4.6 &    23 Oct.   \\
\hline 
HAGN & Gamit-NMA-CML & 12.49 +/-  0.62 &   4.38 +/-   0.68 &  -78.7 +/-    8.8 &   12 Oct.   \\
     & GipsyX-FID-CML & 12.39 +/-  0.50 &   4.34 +/-   0.67 &  -61.6 +/-    8.8 &   29 Oct.   \\
     & GipsyX-NNR-CML & 12.08 +/-  0.54 &   4.42 +/-   0.70 &  -61.7 +/-    9.0 &   29 Oct.   \\
     &     GNSS-CML & 11.95 +/-  0.56 &   4.28 +/-   0.65 &  -50.2 +/-    8.6 &   10 Nov.   \\
\hline 
LYRS & Gamit-NMA-CML &  7.80 +/-  0.35 &   3.09 +/-   0.37 &  -53.4 +/-    6.8 &   6 Nov.   \\
     & GipsyX-FID-CML &  8.26 +/-  0.37 &   3.00 +/-   0.41 &  -67.1 +/-    7.8 &   23 Oct.   \\
     & GipsyX-NNR-CML &  8.20 +/-  0.57 &   3.87 +/-   0.60 &  -99.3 +/-    8.9 &  21 Sep.   \\
     &     GNSS-CML &  8.16 +/-  0.42 &   3.21 +/-   0.45 &  -80.8 +/-    8.0 &  10 Oct.   \\
\hline 
SVES & Gamit-NMA-CML &  6.49 +/-  0.43 &   3.95 +/-   0.45 &  -80.6 +/-    6.5 &  10 Oct.   \\
     & GipsyX-FID-CML &  6.21 +/-  0.45 &   3.46 +/-   0.47 & -103.5 +/-    7.8 &  17 Sep.   \\
     & GipsyX-NNR-CML &  6.27 +/-  0.46 &   3.59 +/-   0.48 & -102.4 +/-    7.6 &   18 Sep.   \\
     &     GNSS-CML &  6.21 +/-  0.45 &   3.37 +/-   0.47 &  -96.8 +/-    7.9 &   23 Sep.   \\
\hline 
HORN & Gamit-NMA-CML &  9.46 +/-  0.27 &   3.40 +/-   0.30 &  -55.4 +/-    5.0 &    4 Nov.   \\
     & GipsyX-FID-CML &  9.52 +/-  0.25 &   3.58 +/-   0.28 &  -62.6 +/-    4.4 &   28 Oct.   \\
     & GipsyX-NNR-CML &  9.59 +/-  0.26 &   3.66 +/-   0.29 &  -63.8 +/-    4.5 &   27 Oct.   \\
     &     GNSS-CML &  9.45 +/-  0.27 &   3.21 +/-   0.30 &  -60.0 +/-    5.3 &    31 Oct.   \\
\hline 
\end{tabular}}
\end{table*}

\begin{table*}
\caption{Yearly amplitude and RMS in the time series. In each column the three parameters are amplitude of yearly signal in mm, \RMS\ of the time series in mm and changes in \RMS\ relative to the unfiltered time series in percent. The numbers are in mm. $H_{GNSS}$, $L_i$ and $H_{CM,L_i}$ are explained in the text.} 
\label{tab:sval-quality-detailed}
\small{
\centering
\begin{tabular}{lrrrrrr} \hline
  Station & $H_{GNSS}$ & $H_{GNSS}-L_1$ &  $H_{GNSS}-L_2$& $H_{CM}$ & $H_{CM,L_1}$ &  $H_{CM,L_2}$ \\ \hline
\hline BJOS\\ Gamit\_NMA&  3.3 / 4.3(   0\%)&  4.9 / 4.3(  -0\%)&  5.0 / 4.3(  -1\%) \\
 GipsyX\_NNR&  1.6 / 5.0(   0\%)&  3.3 / 4.2( -15\%)&  3.4 / 4.2( -15\%) \\
 GipsyX\_FID&  0.9 / 5.2(   0\%)&  2.2 / 4.3( -17\%)&  2.3 / 4.3( -17\%) \\
 GNSS&  1.3 / 4.9(   0\%)&  2.9 / 4.1( -16\%)&  3.0 / 4.1( -16\%) \\
\hline NYA1\\ Gamit\_NMA&  5.8 / 4.5(   0\%)&  6.0 / 4.6(   3\%)&  5.8 / 4.5(   0\%)&  4.1 / 4.3(  -5\%)&  3.1 / 4.3(  -4\%)&  2.0 / 4.2(  -6\%) \\
 GipsyX\_NNR&  3.0 / 5.1(   0\%)&  3.0 / 4.4( -14\%)&  3.3 / 4.2( -16\%)&  2.9 / 5.0(  -2\%)&  2.0 / 4.6(  -8\%)&  0.8 / 4.6( -10\%) \\
 GipsyX\_FID&  3.0 / 5.0(   0\%)&  2.3 / 4.2( -16\%)&  2.4 / 4.1( -18\%)&  2.8 / 5.0(   0\%)&  1.9 / 4.8(  -5\%)&  0.7 / 4.7(  -6\%) \\
 GNSS&  3.8 / 4.6(   0\%)&  3.7 / 4.0( -13\%)&  3.7 / 3.9( -16\%)&  3.4 / 4.7(   2\%)&  2.5 / 4.5(  -3\%)&  1.5 / 4.4(  -4\%) \\
\hline NYAL\\ Gamit\_NMA&  5.2 / 4.7(   0\%)&  5.5 / 4.8(   1\%)&  5.3 / 4.7(  -1\%)&  3.6 / 4.2( -11\%)&  2.7 / 4.1( -13\%)&  1.4 / 4.1( -14\%) \\
 GipsyX\_NNR&  3.2 / 5.1(   0\%)&  2.7 / 4.4( -13\%)&  2.7 / 4.3( -16\%)&  3.5 / 5.1(   0\%)&  2.5 / 4.7(  -7\%)&  1.2 / 4.6(  -9\%) \\
 GipsyX\_FID&  3.4 / 4.9(   0\%)&  2.0 / 4.2( -15\%)&  1.7 / 4.0( -17\%)&  3.3 / 5.0(   3\%)&  2.3 / 4.7(  -4\%)&  1.1 / 4.6(  -6\%) \\
 GNSS&  3.6 / 4.6(   0\%)&  3.1 / 4.1( -12\%)&  3.0 / 3.9( -15\%)&  3.4 / 4.7(   3\%)&  2.5 / 4.5(  -3\%)&  1.3 / 4.4(  -4\%) \\
\hline LYRS\\ Gamit\_NMA&  5.3 / 4.8(   0\%)&  6.2 / 4.9(   3\%)&  6.4 / 4.9(   1\%)&  3.1 / 4.4(  -8\%)&  2.3 / 4.3( -11\%)&  1.6 / 4.2( -12\%) \\
 GipsyX\_NNR&  2.5 / 7.8(   0\%)&  1.6 / 7.3(  -7\%)&  2.4 / 7.2(  -8\%)&  3.4 / 7.8(   0\%)&  2.2 / 7.6(  -4\%)&  1.4 / 7.5(  -4\%) \\
 GipsyX\_FID&  3.1 / 7.0(   0\%)&  2.6 / 6.5(  -7\%)&  2.9 / 6.4(  -8\%)&  2.9 / 7.0(   0\%)&  1.8 / 6.9(  -2\%)&  0.9 / 6.8(  -3\%) \\
 GNSS&  3.0 / 6.8(   0\%)&  2.4 / 6.3(  -8\%)&  2.9 / 6.2( -10\%)&  2.9 / 6.7(  -1\%)&  1.7 / 6.6(  -3\%)&  0.6 / 6.6(  -4\%) \\
\hline SVES\\ Gamit\_NMA&  4.6 / 5.7(   0\%)&  4.8 / 5.7(   1\%)&  4.9 / 5.6(  -1\%)&  3.6 / 5.2(  -7\%)&  2.3 / 5.1( -10\%)&  0.7 / 4.9( -12\%) \\
 GipsyX\_NNR&  2.2 / 6.9(   0\%)&  1.8 / 6.3(  -9\%)&  3.0 / 6.2( -11\%)&  3.0 / 6.8(  -2\%)&  1.8 / 6.5(  -7\%)&  1.3 / 6.4(  -8\%) \\
 GipsyX\_FID&  2.7 / 6.9(   0\%)&  0.9 / 6.2( -11\%)&  2.2 / 6.1( -12\%)&  2.9 / 6.8(  -2\%)&  1.7 / 6.5(  -7\%)&  1.4 / 6.4(  -8\%) \\
 GNSS&  2.5 / 6.5(   0\%)&  1.7 / 5.8(  -9\%)&  2.8 / 5.7( -11\%)&  2.9 / 6.3(  -2\%)&  1.5 / 6.1(  -5\%)&  1.1 / 6.0(  -7\%) \\
\hline HORN\\ Gamit\_NMA&  5.6 / 4.9(   0\%)&  6.0 / 5.1(   3\%)&  5.9 / 4.9(  -0\%)&  3.3 / 4.5(  -8\%)&  2.7 / 4.4( -10\%)&  1.4 / 4.3( -14\%) \\
 GipsyX\_NNR&  3.4 / 4.8(   0\%)&  3.4 / 4.1( -14\%)&  3.4 / 3.9( -19\%)&  3.3 / 4.8(   1\%)&  2.9 / 4.5(  -5\%)&  1.2 / 4.3( -10\%) \\
 GipsyX\_FID&  3.5 / 4.8(   0\%)&  2.9 / 4.0( -17\%)&  2.5 / 3.8( -21\%)&  3.3 / 4.9(   1\%)&  2.8 / 4.7(  -3\%)&  1.2 / 4.4(  -8\%) \\
 GNSS&  3.3 / 4.7(   0\%)&  3.3 / 4.1( -14\%)&  3.4 / 3.8( -20\%)&  3.0 / 4.7(  -0\%)&  2.4 / 4.5(  -5\%)&  1.0 / 4.3(  -9\%) \\
\hline
\end{tabular}}
\end{table*}

\end{document}